\documentclass[aps, prd,english,superscriptaddress,preprintnumbers, onecolumn,floatfix,nofootinbib]{revtex4-1}
\usepackage{amsmath,amssymb,amsfonts,mathrsfs,graphicx,tabularx,url,multirow,hyperref,slashed}
\usepackage[T1]{fontenc}
\usepackage{latexsym,epsfig}
\usepackage[latin9]{inputenc}
\usepackage{babel}

\makeatletter
\@ifundefined{textcolor}{}
{%
 \definecolor{BLACK}{gray}{0}
 \definecolor{WHITE}{gray}{1}
 \definecolor{RED}{rgb}{1,0,0}
 \definecolor{GREEN}{rgb}{0,1,0}
 \definecolor{BLUE}{rgb}{0,0,1}
 \definecolor{CYAN}{cmyk}{1,0,0,0}
 \definecolor{MAGENTA}{cmyk}{0,1,0,0}
 \definecolor{YELLOW}{cmyk}{0,0,1,0}
}

\def\etmiss {{\not\!\! E_T}}

\makeatother
\begin{document}

\title{Heavy Higgs Coupled to Vector-like Quarks: Strong CP Problem and\\
Search Prospects at the 14 TeV LHC}

\author{Alexandre Alves}
\email{aalves@unifesp.br}
\affiliation{Departamento de Ci\^encias Exatas e da Terra,\\
Universidade Federal de S\~ao Paulo, 09972-270, Diadema-SP, Brasil}

\author{Daniel A. Camargo}
\email{dacamargov@gmail.com}
\author{Alex G. Dias}
\email{alex.dias@ufabc.edu.br}
\affiliation{Centro de Ciências Naturais e Humanas,\\
Universidade Federal do ABC, 09210-580, Santo André-SP, Brasil.}

\begin{abstract}
Motivated by a solution to the strong CP problem we propose a model where a new heavy neutral CP-even Higgs boson couples to vector-like quarks enhancing its production cross section whose dominant decays are into weak bosons. The masses of the vector-like quarks are generated through interactions with a singlet scalar field charged under a broken global $U(1)$ symmetry providing a solution to the strong CP problem by means of the Peccei-Quinn mechanism. The diboson excess observed by the ATLAS Collaboration is discussed as the new heavy Higgs boson is a candidate to explain a possible signal in this channel. We also show that the 14 TeV LHC is capable of discovering this heavy Higgs with masses up to 1 TeV in the $H\to ZZ\to \ell^+\ell^-\ell^{\prime +}\ell^{\prime -}$ search channel using boosted decision trees to better discriminate between signals and backgrounds and to tame systematic uncertainties in the background rates.
\end{abstract}

\maketitle

\section{Introduction}

The present data on the Higgs boson, showing that its properties are close to the ones
predicted by the Standard Model (SM), are a landmark that has brought us a decisive
understanding of the spontaneous symmetry breaking mechanism and the interactions of the elementary particles. 
Observations from the ATLAS and CMS Collaborations, on the production
cross section and the main decay rates of the
Higgs boson~\cite{Aad:2012tfa,Chatrchyan:2012xdj,pas-atlas-cms},
also reveal that this particle must have suppressed couplings, in comparison to its
coupling with the top quark, with any hypothetical extra colored fermion. The reason is
that, in the absence of a cancellation mechanism, such colored fermions might contribute
to the Higgs boson production cross section.

An important question to be addressed is whether other fundamental particles
play a role in the mechanism of electroweak symmetry breaking, along with the Higgs boson.
One of the simplest extensions of the SM containing additional Higgs bosons that may be
involved in such symmetry breaking is the two Higgs doublet model (2HDM).
A recent review of 2HDM can be found in Ref.~\cite{Branco:2011iw}. A common feature of any
2HDM is that its particle spectrum contains two CP even neutral scalars, a light one
denoted by $h$ and a heavy one denoted by $H$.
As it happens, $h$ is identified with the discovered Higgs boson, whose mass is approximately $125$ GeV,
while $H$ is expected to have mass well above this value.

We consider a model that, besides the SM fermionic field content, has two Higgs doublets,
a scalar singlet field, and vector-like quarks disposed in $SU(2)_L$ doublet and
singlet representations with hypercharges 7/6 and 5/3, respectively.
The model has a global $U(1)_{PQ}$ anomalous symmetry making up a solution to the strong CP
problem through the Peccei-Quinn mechanism~\cite{Peccei:1977hh}.
The spontaneous breaking of $U(1)_{PQ}$  at a very high energy scale
leads to a light pseudo Nambu-Goldstone boson field whose particle excitation is the
axion~\cite{Peccei:1977hh,Weinberg:1977ma,Wilczek:1977pj}. Our construction can be viewed
as an hybrid axion model containing vector-like singlet quarks, like in the  KSVZ
model~\cite{Kim:1979if},~\cite{Shifman:1979if}, and two Higgs doublets, like in the
DFSZ model~\cite{Dine:1981rt},~\cite{Zhitnitsky:1980tq}.
As in the KSVZ and DFSZ models, the axion in our model is also a dark matter candidate
(for a review of axions and strong CP problem,
see~\cite{Kim:2008hd,Jaeckel:2010ni,Ringwald:2012hr} and references therein).
Another peculiar feature of the model is that masses above hundreds of GeV for the vector-like
quarks are naturally generated through effective operators, suppressed by the
Planck scale, when a scalar singlet field gets a vacuum expectation value (VEV)
breaking the $U(1)_{PQ}$ symmetry. The low energy effective theory resulting
from this symmetry breakdown is a type-I 2HDM in which one Higgs doublet couples to the fermions of the SM, while the other one couples to the vector-like quarks.

The scenario that we take into account in our analysis is one in which $H$ couples mostly to the
vector-like $X$-quarks and, concomitantly, has negligible couplings with the SM
fermions. This is done in order to make the couplings and production
cross section of $h$ compatible with the observations. Therefore, the production
of $H$ by gluon fusion has dominant contribution from vector-like quarks.
Also, in the considered scenario the dominant decay channels of $H$ are the ones
involving a pair of massive vector bosons, that is, $H\rightarrow ZZ$ and
$H\rightarrow W^+W^-$\footnote{A study of the process $H\rightarrow W^+\,W^-$ in a 2HDM containing
vector-like leptons was done in Ref.~\cite{Dermisek:2015oja}}. In our analysis, we took into account the latest
experimental limits on the production of a heavy Higgs boson decaying into  several final states~\cite{Khachatryan:2015yea,Aad:2014yja,Khachatryan:2015cwa,Aad:2014ioa,Khachatryan:2015qba}.

We also study in this work the possibility of having more vector-like
quarks coupled to $H$, and so enhancing its production cross section, in order
to explain the ATLAS Collaboration observation of an excess in the pair production channels $WZ$, $WW$, and $ZZ$
decaying into jets with an invariant mass in the interval
1.8 TeV -- 2 TeV~\cite{Aad:2015owa}.  Some works have proposed that a scalar 2 TeV resonance is responsible for the reported
excess~\cite{Aguilar-Saavedra:2015rna,Arnan:2015csa,Sanz:2015zha,Chiang:2015lqa,Cacciapaglia:2015nga,
Omura:2015nwa,Chao:2015eea,Petersson:2015rza,Zheng:2015dua,Fichet:2015yia,Chen:2015cfa,Sierra:2015zma,Allanach:2015blv}\footnote{There are also many other alternatives involving vector resonances, see Refs.~\cite{Hisano:2015gna,Cheung:2015nha,Dobrescu:2015qna,Gao:2015irw,Cao:2015lia} for example.}.
Two of them specifically present a construction based on a new neutral heavy Higgs coupling
to heavy vector-like quarks. In Ref.~\cite{Chen:2015cfa}, a new SM singlet Higgs couples
to many new quark triplets in order to raise the Higgs production cross section and explain
the excess in the $ZZ$ and $WW$ channels. In Ref.~\cite{Sierra:2015zma}, on the other hand,
the heavy Higgs results from a type-I 2HDM coupling with vector-like quarks
in higher $SU(3)_C$ representations to increase the production cross section of the heavy Higgs. We show that, in order to explain the diboson excess, we need too many new quarks which might raise issues concerning the stability of the scalar potential and spoil the asymptotic freedom of the QCD. On the other hand, we should mention that new data from the 13 TeV LHC do not corroborate the excess found in run I.

It is also important to mention that several models with vector-like quarks and leptons as well as models with axion-like particles decaying to photons have received a lot of attention by virtue of the excess in diphoton events observed by both ATLAS~\cite{atlas} and CMS~\cite{cms}. In this respect, as we are going to show, the model presents a pseudoscalar Higgs boson whose production cross section could be enhanced by the vector-like quarks and which could explain that excess. We do not pursue this investigation, but postpone it for a future work.

In an environment of low signal to background ratio, systematic uncertainties might not allow the high significance necessary for discovery at the LHC. Cut-based analysis often do not help to raise that ratio and multivariate analysis are becoming essential to hunt for feeble new physics signals. One of those techniques is Boosted Decision Trees (BDT), a supervised machine learning algorithm that has been successfully employed in high energy physics. We will show that using BDT, broad resonances corresponding to the decay of heavy Higgses with masses up to 1 TeV can be reached at the 14 TeV LHC, in the golden-channel $H\to ZZ\to \ell^+\ell^-\ell^{\prime +}\ell^{\prime -}\, ,\, \ell(\ell^\prime)=e,\mu$, with ${\cal O}$(20\%) systematics in the background rate. On the other hand, a cut-based analysis fails to deliver a $5\sigma$ signal even with rather low systematics. As far as we know, this is the first study where a machine learning technique is used to classify signal events of a broad resonance where a cut on the invariant mass is not effective.

We want to emphasize that our work is different from the Refs.~\cite{Chen:2015cfa,Sierra:2015zma} in the sense of its very own motivation -- we construct a model, a Peccei-Quinn symmetry implemented with a doublet plus one singlet of vector-like quarks, aimed essentially to solve the strong CP problem. 

In Sec.~\ref{sec:scp}  we introduce the model; the results of our
phenomenological analysis are presented in Sec.~\ref{sec:pheno} followed
by the conclusions in Sec.~\ref{sec:conclusions}.

\section{A two Higgs Doublet Model with Vector-like quarks and Free from the Strong CP Problem}
\label{sec:scp}

For the field content of the model we have as scalar fields a complex singlet plus two doublets
\begin{equation}
S\sim \left(\mathbf{1},\,0\right),\hspace{0.8 cm}
\Phi_{a}=\left[\begin{array}{c}\phi_a^+\\
\phi_a^0\end{array}\right]\sim\left(\mathbf{2},\,1/2\right),\hspace{0.3 cm}a=1,2;
\label{sdfds}
\end{equation}
where inside the parenthesis are the quantum numbers under $SU(2)_L$ and the
hypercharge $\mathrm{U}(1)_Y$, respectively. It is convenient to parametrize the complex
scalar singlet as $S=\rho(x)e^{i\frac{a(x)}{f_a}}/\sqrt2$, where $a(x)$ is the axion field
and $f_a$ its decay constant.
In addition to the SM fermionic fields we
include a doublet and a singlet of vector-like quarks
\begin{equation}
\psi_{L,R}^{X}=\left[\begin{array}{c}
X_{L,R}^{\prime}\\
U_{L,R}^{\prime}
\end{array}\right]\sim\left(\mathbf{2},\,7/6\right),
\quad X_{L,R}^{\prime\prime}\sim\left(\mathbf{1},\,5/3\right).
\label{qvmult}
\end{equation}
When denoting these quarks as \emph{vector-like quarks} we are referring to their properties under
the SM $SU(2)_L\otimes U(1)_Y$ gauge group. In fact, such quarks are chiral under the $U(1)_{PQ}$ symmetry.
The fermionic multiplets in Eq. (\ref{qvmult}) compose the minimal setup of this model.
The heavy Higgs production cross section depends on the number of vector-like quarks as
well as their masses. Thus, we also consider in our phenomenological analysis a
generalization in which the set of vector-like quarks multiplets in Eq. (\ref{qvmult})
is replicated so that the number $N_X$ of $X$-type quark fields can be such that
$N_X \geq2$ ($N_X=2$ for the one set case).

We found that the composition of a doublet plus a singlet of vector-like quarks in
Eq. (\ref{qvmult}) is one of the minimal set of quarks allowing an implementation of the
$U(1)_{PQ}$ symmetry and, at the same time,  coupling to a Higgs doublet from which a heavy Higgs boson is originated. Such a heavy Higgs boson has its production through gluon fusion enhanced due the couplings with the vector-like quarks.  

We emphasize that vector-like quarks in the representations which we take into account
here were considered separately in different studies constraining their masses and mixing with
SM fermions~\cite{AguilarSaavedra:2009es,Okada:2012gy,
Cacciapaglia:2011fx,Cacciapaglia:2012dd,Aguilar-Saavedra:2013qpa,
Barducci:2014ila,Cacciapaglia:2015ixa,Alok:2015iha,Cao:2015doa}.

The fields in Eqs. (\ref{sdfds}) and (\ref{qvmult}) are assumed to carry charge of $U(1)_{PQ}$ symmetry
as shown in Table \ref{tableupq}, with all SM fields not carrying charge of this symmetry.
It can be seen from the quarks $U(1)_{PQ}$ charges in Table \ref{tableupq} that a nonzero value
for the anomaly coefficient is obtained, that is,
\begin{equation}
C_{ag}=2(X_{\psi L}-X_{\psi R})+X_{X_{L}^{\prime\prime}}-X_{X_{R}^{\prime\prime}}=2\, .
\label{agcoeff}
\end{equation}
Thus, the axion field, $a(x)$, couples to the gluon's field strength,
$G^b_{\mu\nu}$, as required for solving the strong CP problem through the
Peccei-Quinn mechanism: the CP violating parameter
$\bar{\theta}$ is replaced by  $a(x)$ in the Lagrangian term
$\mathscr{L}\supset\bar{\theta}G^b_{\mu\nu}\tilde{G}^{b,\mu\nu}$, in which
$\tilde{G}^{b,\mu\nu}=\epsilon^{\mu\nu\rho\sigma}G^b_{\rho\sigma}/2$; and the potential
$V(a(x))$ has a minimum such that CP is conserved in the strong interactions.

\begin{table}[h]
\[
\begin{array}{c|ccccccc}
\hline
\Psi  & S & \Phi_1 & \Phi_2 & \psi_{L}^{X} & \psi_{R}^{X} & X_{L}^{\prime\prime}
& X_{R}^{\prime\prime}  \\
\hline
X_\Psi & 1 & 1  & 0 & 1 & -1 & 0 & 2
\\[.5ex]
\hline
\end{array}
\]
\caption{\label{tableupq} $U(1)_{PQ}$ charges, $X_\Psi$, defined in the field's
transformations $\Psi\rightarrow e^{i\alpha X_\Psi}\Psi $. The SM fermionic fields do not carry charge of U(1)$_{PQ}$ and are not shown. }
\end{table}

The scalar potential containing renormalizable operators invariant under $U(1)_{PQ}$ is
\begin{align}
V(S,\Phi_{a}) & =-\mu^2_S \mid S\mid^2-\mu_{11}^{2}\mid\Phi_{1}\mid^{2}
-\mu_{22}^{2}\mid\Phi_{2}\mid^{2}
-\mu_{12}\left(S\,\Phi_{1}^{\dagger}\Phi_{2} +S^\dagger\Phi_{2}^{\dagger}\Phi_{1}\right)
\nonumber \\
& +\left(\frac{\lambda_{S}}{2}\mid S\mid^{2}+\lambda_{1S}\mid\Phi_{1}\mid^{2}
+\lambda_{2S}\mid\Phi_{2}\mid^{2}\right)\mid S\mid^2+\frac{\lambda_{1}}{2}\mid\Phi_{1}\mid^{4}
+\frac{\lambda_{2}}{2}\mid\Phi_{2}\mid^{4}\nonumber \\
& +\lambda_{3}\mid\Phi_{1}\mid^{2}\mid\Phi_{2}\mid^{2}
+\lambda_{4}\mid\Phi_{1}^{\dagger}\Phi_{2}\mid^{2}\,,
\label{potupq}
\end{align}
where all the parameters are taken as being real.

Yukawa interactions respecting the $U(1)_{PQ}$ symmetry are such that SM
left-handed doublets of quarks and leptons $q_{i}$, $L_i$, and right-handed singlets  of quarks $u^{\prime}_{iR}$,
$d^{\prime}_{iR}$, and leptons, $l^{\prime}_{iR}$, weak eigenstates couples only to $\Phi_2$ according to the terms
\begin{equation}
\mathscr{L}\supset -y_{ij}^{u}\,\overline{q_i}\widetilde{\Phi}_{2}u_{jR}^{\prime}
-y_{ij}^{d}\,
\overline{q_i}{\Phi}_{2}d_{jR}^{\prime}-y_{ij}^{l}\,
\overline{L_i}{\Phi}_{2}l_{jR}^{\prime}+h.c.\,,
\label{yukqsm}
\end{equation}
with $i=1,2,3$ the generation index, and the vector-like quarks couples  $\Phi_{1}$ according to the terms
\begin{equation}
\mathscr{L}\supset -y_{i}^{u}\overline{\psi_{L}^{X}}\Phi_{1}u_{iR}^\prime-y^\prime
\overline{\psi_{L}^{X}}\widetilde{\Phi}_{1}X_{R}^{\prime\prime}
-y^{\prime\prime}\overline{X_{L}^{\prime\prime}}\widetilde{\Phi}_{1}^{\dagger}\psi_{R}^{X}
+h.c
\label{yukqx}
\end{equation}
with $\widetilde{\Phi}_{1,2}=\epsilon\Phi_{1,2}^*$, and $y_{ij}^u,y_{ij}^d,y_{ij}^l,y_{i}^u,y^\prime,y^{\prime\prime}$ denoting the Yukawa coupling constants. The first term in Eq. (\ref{yukqx})
is the main interaction from which the vectorial quarks may decay into SM particles.
Additionally, we consider the interactions of vector-like quarks with $S$ through effective
operators suppressed by a high energy scale $\Lambda$,
\begin{equation}
\mathcal{L}\supset\frac{g_{S}^\prime}{\Lambda}
S^{2}\overline{\psi_{L}^{X}}\psi_{R}^{X}+\frac{g_{S}^{\prime\prime}}{\Lambda}
S^{*2}\overline{X_{L}^{\prime\prime}}X_{R}^{\prime\prime}+h.c.
\label{yukpx}
\end{equation}

Spontaneous breakdown of the $U(1)_{PQ}$ symmetry is realized through the VEV of the scalar singlet
$\langle S\rangle=V_S/\sqrt2$, which is assumed to happen at a energy scale much above the one where the gauge symmetry $SU(2)_L\otimes U(1)_Y$ is broken by the VEVs of
$\langle \Phi_{1,2}\rangle=[0,\,v_{1,2}/\sqrt2]^T$, i. e., $V_S\gg v$, with
$v=\sqrt{v_1^2+v_2^2}=246$ GeV. This assumption makes the axion a very light and weak
interacting particle since both its mass and couplings are suppressed by the decay
constant $f_a$, which is approximately  $f_a\approx V_S$ (see, for example,~\cite{Dias:2014osa}).

The VEV $\langle S\rangle=V_S/\sqrt2$ leads to a generation of mass terms for the vector-like
quarks as
\begin{equation}
\mathcal{L}\supset M^\prime\overline{\psi_{L}^{X}}\psi_{R}^{X}
+M^{\prime\prime}\overline{X_{L}^{\prime\prime}}X_{R}^{\prime\prime}+h.c.
\label{yukpx}
\end{equation}
in which
\begin{equation}
M^\prime=\frac{g_{S}^\prime V_S^2}{\Lambda},\hspace{0.8 cm}
M^{\prime\prime}=\frac{g_{S}^{\prime\prime} V_S^2}{\Lambda}.
\label{mlmll}
\end{equation}
There will be still contributions proportional to $v_1$, due the VEV
$\langle \Phi_1\rangle$, to the masses of the vector-like quarks, but we assume
that  $M^{\prime}$ and $M^{\prime\prime}$ are the main contributions.
With these masses within a interval of hundreds of GeV and few TeV, the vector-like
quarks might leave observable signals of physics beyond the SM at the LHC like in
the production process of the heavy
Higgs boson, which is one of our aims in this work. For example, taking the value
$f_a\approx V_S\approx10^{11}$ GeV the axion a mass and its coupling with two photons
have the values
\begin{eqnarray}
& & m_a=\frac{m_\pi f_\pi}{f_a}\frac{\sqrt z}{1+z}\approx 6\times10^{-5}\,{\rm eV};
\nonumber \\
& & g_{a\gamma\gamma}\approx \frac{\alpha}{2\pi f_a}\left(\frac{C_{a\gamma}}{C_{ag}}
-\frac{2}{3}\frac{4+z}{1+z}\right)
\approx8.3\times 10^{-15}\,{\rm GeV^{-1}}.
\label{max}
\end{eqnarray}
where $m_\pi=135$ MeV is the pion mass, $f_\pi=92$ MeV the pion decay constant,
$z=m_u/m_d\approx 0.56$ the mass ratio of the up and down quarks, and the axion-photon
anomaly coefficient for this model is $C_{a\gamma}=16/3$. With the parameter  values
in Eq. (\ref{max}), axions can constitute a relevant component of dark matter in the
Universe\footnote{Effective operators suppressed by the Planck scale could potentially
destabilize the dark candidates supposedly made stable by an exact  global continuous
symmetry~\cite{Mambrini:2015sia}. The axion is free from this problem because its stability
does not rely such a symmetry. Also, it can be seen that gravity induced operators
violating $U(1)_{PQ}$ like, for example, $S\,F^{\mu\nu}F_{\nu\mu}/M_{Pl}$, do not lead to
a time life for the axion, as ultra-light dark matter candidate, which is lower than
the age of the Universe.}.
See Ref.~\cite{Dias:2014osa} for the actual limits and allowed parameter space for the axion.
Thus, for  $\Lambda=2.4\times10^{18}$ GeV (the reduced Planck scale) we have that
$M^\prime$, $M^{\prime\prime}\sim1$ TeV if $g_{S}^{\prime},\,g_{S}^{\prime\prime}\simeq0.25$.
We then see that TeV masses to the vector-like quarks can be originated in this scheme.

One observation concerning the axion models is that semi-classical gravity effects
might not preserve global symmetries and this prevent the solution to the
strong CP problem, once the  $U(1)_{PQ}$ symmetry would be also explicitly  violated
by Planck scale suppressed nonrenormalizable operators of the form
${\cal O}\sim S^D/M_{Pl}^{D-4}$ leading to dangerous corrections to the axion
potential~\cite{Georgi:1981pu,Ghigna:1992iv,Holman:1992us,Kamionkowski:1992mf,Barr:1992qq}.
A way out of this problem can be achieved through the imposition of certain discrete
$Z_N$ symmetries to forbid  those undesired nonrenormalizable
operators~\cite{Dias:2002hz,Carpenter:2009zs,Harigaya:2013vja,Dias:2014osa}.
An treatment of this question in the present model will be done elsewhere.
We simply suppose here that there is a discrete symmetry such that the $U(1)_{PQ}$ symmetry
is only broken in nonrenormalizable operators with dimension sufficiently high
so that the solution to the strong CP problem is not spoiled.

We suppose that, after spontaneous breaking of $U(1)_{PQ}$, the low energy effective theory contains the vector-like quarks and two Higgs doublets fields. The radial component $\rho(x)$ of $S(x)$ gets a mass $m_\rho= \sqrt{\lambda_S} V_S$. It is assumed that $\lambda_S \sim{\cal O}(1)$ in order that  $m_\rho$ is larger than the masses of the vectorial quarks and the scalars from the doublets. After the integration of $\rho(x)$ the potential, up to quartic terms, turns out to be, effectively, 
\begin{align}
V_{eff}(\Phi_{a}) & \simeq -m_{11}^{2}\mid\Phi_{1}\mid^{2}-m_{22}^{2}\mid\Phi_{2}\mid^{2} -m_{12}^{2}\left(\Phi_{1}^{\dagger}\Phi_{2}+\Phi_{2}^{\dagger}\Phi_{1}\right)
\nonumber \\
& +\frac{\lambda_{1}}{2}\mid\Phi_{1}\mid^{4}+\frac{\lambda_{2}}{2}\mid\Phi_{2}\mid^{4} +\lambda_{3}\mid\Phi_{1}\mid^{2}\mid\Phi_{2}\mid^{2}+(\lambda_{4}+\delta\lambda_4)\mid\Phi_{1}^{\dagger}\Phi_{2}\mid^{2}\nonumber \\
& - \delta\lambda_{12}\left((\Phi_{1}^{\dagger}\Phi_{2})^2+(\Phi_{2}^{\dagger}\Phi_{1})^2\right)+\left(\delta\lambda_{1S}\mid\Phi_{1}\mid^{2} +\delta\lambda_{2S}\mid\Phi_{2}\mid^{2}\right)\left(\Phi_{1}^{\dagger}\Phi_{2}\Phi_{2}^{\dagger}\Phi_{1}\right)
\label{poteff}
\end{align}
with the definition of the parameters as 
\begin{align}
& m_{11}^{2}=\mu_{11}^2-\lambda_{1S}\frac{V_S^2}{2},\hspace{0.6 cm}
m_{22}^{2}=\mu_{22}^2-\lambda_{2S}\frac{V_S^2}{2},\hspace{0.6 cm}
m_{12}^{2}=\mu_{12}\frac{V_S}{\sqrt2} \nonumber \\
& \delta\lambda_4 = -\frac{\mu_{12}^2}{m_\rho^2},\hspace{0.6 cm} 
\delta\lambda_{12} = -\frac{\mu_{12}^2}{2m_\rho^2},\hspace{0.6 cm} 
\delta\lambda_{1S} = \lambda_{1S}\frac{\mu_{12}V_S}{\sqrt{2}m_\rho^2},\hspace{0.6 cm}
\delta\lambda_{2S} = \lambda_{2S}\frac{\mu_{12}V_S}{\sqrt{2}m_\rho^2}.
\label{pareff}
\end{align}
For $\mid\mu_{12}\mid\ll m_\rho$  all quartic terms which arise from the integration of $\rho(x)$ can be disregarded and, thus, we take $\mid\delta\lambda_4\mid \approx \mid\delta\lambda_{12}\mid \approx \mid\delta\lambda_{1S}\mid \approx \mid\delta\lambda_{2S}\mid \approx 0$.  The parameters in  Eq. (\ref{poteff}) are then assumed to be in a region of values where the minimum value of the potential occurs for nonzero VEVs $\langle\Phi_{a}\rangle=[0\,\,v_a/\sqrt{2}]^T$, breaking the $SU(2)\otimes U(1)_Y$ symmetry.

With the two Higgs doublets there are eight degrees of freedom -- four of each doublet in Eq. (\ref{sdfds}) -- being three of them Goldstone bosons absorbed by the $W^\pm$ and $Z$ gauge bosons which then turn out to be massive. The remaining five degrees of freedom compose the set of scalar fields: a charged scalar, $h^\pm$, a pseudoscalar, $A$, plus two scalars $h$ and $H$. The observed Higgs boson having mass approximately 125 GeV is identified here with $h$, while $H$ is the heavy Higgs boson that we will aim in our phenomenological analysis.

In what follows, our analysis will be based on the effective model which has
the Yukawa interactions in Eqs. (\ref{yukqsm}), (\ref{yukqx}), the mass terms
in Eq. (\ref{yukpx}), and the potential in Eq. (\ref{poteff}). The model has
a similarity with the known type-I two Higgs doublet model by the fact that
only one scalar doublet -- in this case $\Phi_2$ -- couples with the SM quark
fields.
The other scalar doublet couples with the new quarks which play a major
role in the production of $H$ through gluon fusion as we will see. This idea of
coupling a second scalar doublet with hypothetical quark fields has already
been explored in models containing chiral quarks~\cite{Alves:2013dga},~\cite{Camargo:2015xca}.

\subsubsection{Masses from spontaneous breaking of the $SU(2)_L\otimes U(1)_Y$ symmetry}

Parameterizing the neutral components in Eq. (\ref{sdfds}) as
\begin{equation}
\phi_{1,2}^0=\frac{1}{\sqrt2}\left(v_{1,2}+\phi_{1,2}^{0}(x)+i\eta_{1,2}^{0}(x)\right),
\end{equation}
the squared masses of the fields are obtained from the bilinear terms in
Eq. (\ref{poteff}) are similar to the type-I two Higgs doublets model. The difference
is that the present model does not have a term $\lambda_5((\Phi_1^\dagger\Phi_2)^2+h.c.)$
in the scalar potential. Thus, squared masses for the the pseudoscalar  and the  charged scalar  are given by
\begin{align}
m_{A}^2 & = \frac{m_{12}^{2}}{s_\beta c_\beta}\,,\nonumber \\
m_{h^\pm}^2 & = m_{A}^2-\lambda_{4}v^2\,,
\label{mesc}
\end{align}
while the squared masses of the scalars $h$ and $H$ are given by the eigenvalues
of the mass matrix which is, in the basis
$\left(\,\phi_1^{0}\,\,\,\,\,\phi_2^{0}\,\right)^T$,
\begin{equation}
\left(\begin{array}{cc}
s_\beta^2 m_A^2+\lambda_{1}c_\beta\, v^2 & s_\beta c_\beta\left(
\lambda_{3}\,v^2-m_{h^\pm}^2\right)\\
s_\beta c_\beta\left(
\lambda_{3}\,v^2-m_{h^\pm}^2\right) & c_\beta^2 m_A^2+\lambda_{2}s_\beta\, v^2
\end{array}\right)
\label{mmesc}
\end{equation}
where $s_\beta\,\,(c_\beta)\equiv{\rm sin}\beta\,\,({\rm cos}\beta)$,
$t_\beta=s_\beta/c_\beta\equiv v_2/v_1$. The symmetry
and mass eigenstates are related through an orthogonal transformation,
parametrized by an angle $\alpha$, defined as
\begin{equation}
\left(\begin{array}{c}
\phi_1^{0} \\
\phi_2^{0}
\end{array}\right)=\left(\begin{array}{cc}
-s_\alpha & c_\alpha\\
c_\alpha & s_\alpha
\end{array}\right)
\left(\begin{array}{c}
h \\
H
\end{array}\right)
\label{rhHm}
\end{equation}
We will work in the scenario in which the mixing of these
scalars is small, $s_\alpha\approx 0$, with choices of parameters
such that $|s_\beta c_\beta\left(
\lambda_{3}\,v^2-m_{h^\pm}^2\right)|\ll m_h^2/2$.
It means that the symmetry and mass eigenstates are almost the same, i. e.,
$\phi_2^{0}\approx h$, and $\phi_1^{0}\approx H$, so that $h$ might also have
small Yukawa couplings with the vector-like quarks. This choice is consistent
with the fact that the accumulated  data on the Higgs boson reveal no significant
deviations from the SM predictions.
In this case, from Eq. (\ref{mmesc}), the mass expressions for $h$ and $H$ are
\begin{align}
m_{h}^2 & \approx 2(c_\beta^2 m_A^2+\lambda_{2}s_\beta\, v^2)\,,
\nonumber \\
m_{H}^2 & \approx 2(s_\beta^2 m_A^2+\lambda_{1}c_\beta\, v^2)\,,
\label{mhH}
\end{align}
where $m_h\approx 125$ GeV, while $m_H$ is the mass of the heavy Higgs,
with mass $m_H>m_h$. Observe that for $0\leq\beta\leq\pi/2$, if  $s_\beta^2>1/2$ then $m_H>m_A$, since $\lambda_1>0$ is required to have a stable scalar potential. 

Since the vector-like quarks are coupled with $H_1$ there is a contribution
to their masses due the spontaneous breaking of the $SU(2)_L\otimes U(1)_Y$
symmetry, and proportional to VEV $v_1$. The mass matrix $\mathcal{M}_X$
for the $X'$s quarks is turned into a diagonal form through a bi-unitary
transformation
\begin{equation}
\mathcal{U}_{L}\mathcal{M}_X\,\mathcal{U}_{R}^{\dagger}
=\mathrm{Diag}\left(M_{X1},\, M_{X2}\right),
\label{diagx}
\end{equation}
involving the mixing unitary matrices $\mathcal{U}_{L}$ and $\mathcal{U}_{R}$,
with  $(X_{R}^{0}\,\,X_{R}^{0\prime})$,
\begin{eqnarray}
\mathcal{M}_{X}=\left[\begin{array}{cc}
M^\prime & y^\prime\frac{c_\beta\,v}{\sqrt{2}}\\
y^{\prime\prime}\frac{ c_\beta\,v}{\sqrt{2}} & M^{\prime\prime}
\end{array}\right]=\left[\begin{array}{cc}
M_{0} & m_{L}\\
m_{R} & M_{0}^{\prime}
\end{array}\right]
\end{eqnarray}
defined in the symmetry bases $(X_{L}^{0}\,\,X_{L}^{0\prime})$,
$(X_{R}^{0}\,\,X_{R}^{0\prime})$ and
\begin{equation}
\mathcal{U}_{L}=\left(\begin{array}{cc}
c_{\theta_{L}} & -s_{\theta_{L}}\\
s_{\theta_{L}} & c_{\theta_{L}}
\end{array}\right)\,,
\label{mmx}
\end{equation}
with an equivalent form for $\mathcal{U}_{R}$. We are omitting a complex
phase in Eq. (\ref{mmx}). The mixing angles in $\mathcal{U}_{L}$ and
$\mathcal{U}_{R}$ are defined as
\begin{eqnarray}
t_{2\theta_{L}}=\frac{\sqrt2\,c_\beta\, v (y^\prime M^{\prime\prime}
+y^{\prime\prime} M^\prime)}{2(M^{\prime\prime2}
-M^{\prime2})-(y^{\prime2}-y^{\prime\prime2}) c_\beta^2 v^2},
\label{tanxl}\\
t_{2\theta_{R}}=\frac{\sqrt2\,c_\beta\, v (y^{\prime\prime}\,
M^{\prime\prime}+y^\prime M^\prime)}{2(M^{\prime\prime2}
-M^{\prime2})+(y^{\prime2}-y^{\prime\prime2})c_\beta^2 v^2}.
\label{tanxr}
\end{eqnarray}
Denoting  $X_1$ and $X_2$ the mass eigenstates their respective masses,
from Eq. (\ref{diagx}), are such that
\begin{eqnarray}
M_{X_1}^2 &=&\frac{1}{2}(M^{\prime2}+M^{\prime\prime2})
+\frac{1}{4}(y^{\prime2}+y^{\prime\prime2})c_\beta^2\,v^2\nonumber\\
& &+\frac{1}{2}\sqrt{\left(M^{\prime2}+M^{\prime\prime2}
+\frac{1}{2}(y^{\prime2}+y^{\prime\prime2})c_\beta^2\,v^2\right)^2
-4\left(M^{\prime}M^{\prime\prime}
-\frac{1}{2}y^{\prime}y^{\prime\prime}c_\beta^2\,v^2\right)^2},
\\\nonumber\\
M_{X_2}^2 &=&\frac{1}{2}(M^{\prime2}+M^{\prime\prime2})
+\frac{1}{4}(y^{\prime2}+y^{\prime\prime2})c_\beta^2\,v^2
\nonumber\\
& &-\frac{1}{2}\sqrt{\left(M^{\prime2}+M^{\prime\prime2}+\frac{1}{2}(y^{\prime2}
+y^{\prime\prime2})c_\beta^2\,v^2\right)^2
-4\left(M^{\prime}M^{\prime\prime}-\frac{1}{2}y^{\prime}y^{\prime\prime}
c_\beta^2\,v^2\right)^2}.
\label{tanxr}
\end{eqnarray}

Concerning the  $u_i^\prime-U^\prime$ quarks mass matrix
\begin{eqnarray}
\mathcal{M}_{uU}=\left[\begin{array}{cc}
\left(y^u_{ij}\frac{s_\beta\,v}{\sqrt{2}}\right)_{3\times3} & 0_{3\times1} \\
\left((y^u_{i})^T\frac{c_\beta\,v}{\sqrt{2}}\right)_{1\times3}  & M^{\prime\prime}
\end{array}\right]
\label{mU}
\end{eqnarray}
it could be diagonalized by blocks in the approximation $y^u_{i}\,v\ll M^{\prime\prime}$.
In our analysis we assume $y_{i}^{u}\ll y^{\prime},\,y^{\prime\prime}$ and
$M^{\prime},\,M^{\prime\prime}> v$, so that the mixing among  $u_i$ and
$U$ quarks are sufficiently small to satisfy all actual flavor physics
constraints. It can be seen from Eq. (\ref{mU}) that the Yukawa diagonal
couplings of $H$ with $u_i$ and $U$ quarks mass eigenstates are proportional to
$y^u_i c_\beta\,v /M^{\prime\prime}$. Since we are interested in
the production signals of $H$, whose main process is through gluon fusion,
we disregard such Yukawa diagonal couplings and consider only the
effective ones resulting from  interactions with the $X$ quarks.
Even taken as small the the mixing of $u_i$ and $U$ quarks are the
main interactions for allowing the vectorial quarks decay into SM particles as,
for example, $U\rightarrow d_i\,\,W^+$, and $X \rightarrow u_i\,\,W^+$.

\subsubsection{Couplings of the neutral scalar bosons $h$, $H$, and $A$}

For the trilinear interactions of $h$, $H$ and $A$ with the SM quarks, $q$, and leptons, $l$,
\begin{equation}
\mathscr{L}\supset -\sum_{q} \left(y_{q h}\,h\overline{q} q+y_{q H}\,H\overline{q} q-iy_{q A}\,A\overline{q}\gamma_5 q\right)
- \sum_{l}\left(y_{l h}\,h\overline{l} l+y_{l H}\,H\overline{l} l-iy_{l A}\,A\overline{l}\gamma_5 l\right),
\end{equation}
electroweak vector bosons,
\begin{equation}
\mathscr{L}\supset -\left(g_{VVh}\,h+g_{VVH}H\right)V^{\mu}V_{\mu}\,,
\end{equation}
where $V=W,\,Z$, and vector-like quarks
\begin{equation}
\mathscr{L}\supset -\left(y_{Xh}^{ab} h\overline{X_{a}} X_{b}+y_{XH}^{ab}H\overline{X_{a}} X_{b}-iy_{X A}^{ab}\,A\overline{X_{a}}\gamma_5 X_{b}\right),\,\,\,\,\,\, a,\,b=1,2,
\end{equation}
the effective couplings are: 

for the CP even scalars 
\begin{align}
y_{lh} & = \frac{M_l}{v}\left(\frac{c_{\alpha-\beta}}{t_\beta}-s_{\alpha-\beta}\right)
\,, \hspace{1.7 cm} y_{lH} = \frac{M_l}{v}\left(c_{\alpha-\beta}+\frac{s_{\alpha-\beta}}{t_\beta}\right)
\,, \label{hHyl}\\
y_{qh} & = \frac{M_q}{v}\left(\frac{c_{\alpha-\beta}}{t_\beta}-s_{\alpha-\beta}\right)
\,, \hspace{1.7 cm} y_{qH} = \frac{M_q}{v}\left(c_{\alpha-\beta}+\frac{s_{\alpha-\beta}}{t_\beta}\right)
\,, \label{hHyq}\\
g_{VVh} & = 2\frac{M_V^2}{v}s_{\alpha-\beta}\,, \hspace{3.5 cm}
g_{VVH} = 2\frac{M_V^2}{v}c_{\alpha-\beta}\label{hHvc}\\
y_{X h}^{11} & = -(y^\prime c_{\theta_{L}} s_{\theta_{R}}+ y^{\prime\prime}
s_{\theta_{L}} c_{\theta_{R}})\frac{s_\alpha}{\sqrt{2}} \,,\hspace{1.0 cm}
y_{X H}^{11} = (y^\prime c_{\theta_{L}} s_{\theta_{R}}+ y^{\prime\prime}
s_{\theta_{L}} c_{\theta_{R}})\frac{c_\alpha}{\sqrt{2}}\,,
\label{hHX1}\\
y_{X h}^{22} & = (y^\prime s_{\theta_{L}} c_{\theta_{R}}+ y^{\prime\prime}
c_{\theta_{L}} s_{\theta_{R}})\frac{s_\alpha}{\sqrt{2}} \,,\hspace{1.2 cm}
y_{ X H}^{22} = -(y^\prime s_{\theta_{L}} c_{\theta_{R}}+ y^{\prime\prime}
c_{\theta_{L}} s_{\theta_{R}})\frac{c_\alpha}{\sqrt{2}}\,,
\label{hHX2}\\
y_{X h}^{12} & = (y^\prime c_{\theta_{L}} c_{\theta_{R}}- y^{\prime\prime}
s_{\theta_{L}} s_{\theta_{R}})\frac{s_\alpha}{\sqrt{2}} \,,\hspace{1.2 cm}
y_{ X H}^{12} = (y^\prime c_{\theta_{L}} c_{\theta_{R}}- y^{\prime\prime}
s_{\theta_{L}} s_{\theta_{R}})\frac{c_\alpha}{\sqrt{2}}\,,
\label{hHX12}
\end{align}

and for the CP  odd scalar
\begin{align}
y_{u A} & = \frac{M_u}{v\,t_\beta}\,, \hspace{0.6 cm} y_{d A}  = -\frac{M_d}{v\,t_\beta}\,, \hspace{0.6 cm} y_{l A}  = -\frac{M_l}{v\,t_\beta}\,,
\label{udlA}\\
y_{X A}^{11} & = (y^\prime c_{\theta_{L}} s_{\theta_{R}}- y^{\prime\prime}
s_{\theta_{L}} c_{\theta_{R}})\frac{s_\beta}{\sqrt{2}}\,,
\label{AX1}\\
y_{X A}^{22} & =  -(y^\prime s_{\theta_{L}} c_{\theta_{R}}- y^{\prime\prime}
c_{\theta_{L}} s_{\theta_{R}})\frac{s_\beta}{\sqrt{2}}\,,
\label{AX2}\\
y_{X A}^{12} & =  (y^\prime c_{\theta_{L}} c_{\theta_{R}}+ y^{\prime\prime}
s_{\theta_{L}} s_{\theta_{R}})\frac{s_\beta}{\sqrt{2}}\,,
\label{AX12}
\end{align}

The couplings in Eqs. (\ref{hHyl}), (\ref{hHyq}),  (\ref{hHvc}), and (\ref{udlA}) are the same  of the type-I 2HDM, with $M_l$, $M_q$, and $M_V$ denoting the masses of the SM leptons, quarks,  and electroweak vector bosons, respectively. In the limit
$\alpha\rightarrow 0$, it is observed that $h$ decouples from vector-like
quarks, and that $H$ decouples from SM fermions. But due the diagonal
couplings $y_{X H}^{aa}$  in (\ref{hHX1}) and (\ref{hHX2}), $H$ can
still be produced via gluon fusion process dominantly over the vector
boson fusion for a considerable range of the $X$ quarks masses. 

We observe from Eqs. (\ref{AX1}), (\ref{AX2}), and (\ref{AX12}) that the couplings of $A$ with the $X$ quarks are not suppressed in the limit $\alpha\rightarrow 0$. Thus, in comparison with $H$, we see that $A$ may have a high enough production cross section -- which is essentially given  through gluon fusion process -- to allow its observation at LHC. Another property of $A$ is that it has effective couplings, generated at one-loop order in perturbation theory, with the electroweak vector bosons $W$ and $Z$. Such couplings can be comparable in strength to the one of $A$ to two photons. These features may endow $A$ as a candidate for the recent evidence of a 750 GeV resonance present by ATLAS~\cite{atlas} and CMS~\cite{cms} Collaborations.

\subsubsection{Heavy Higgs-gluon effective Lagrangian }

The effective Lagrangian for the interaction of one $H$ with two gluons can be derived in the same way as the Higgs boson-electromagnetic field effective
Lagrangian~\cite{Ellis:1975ap},~\cite{Shifman:1979eb}. Adapting the
calculations of Vainstein {\it et al}~\cite{Shifman:1979eb} to the case of the  gluon 
field strength, the one-loop corrections
(see Ref.~\cite{Itzykson:1980rh}) due the number $N_X$ of  $X$ quarks are
\begin{equation}
\delta\mathscr{L} =-\frac{1}{4}\sum_{i=1}^{N_X}\frac{g_{s}^2}{24\pi^{2}}
 {\rm ln}\left(\frac{M_{X_i}^{2}}{\Lambda^{2}}\right)\,G^b_{\mu\nu}G^{b,\mu\nu}\,,
\end{equation}
where $\Lambda$ is the energy cutoff. With the replacement
$M_{X_i}\rightarrow M_{X_i}+y^{ii}_{XH}H$  and defining the ratios
$r_i=y^{ii}_{XH} v/M_{X_i}<1$ we get the effective Lagrangian 

\begin{equation}
\mathscr{L}_{HGG}  = -\frac{\alpha_s}{12\pi v}
r H\, G^b_{\mu\nu}G^{b,\mu\nu}.
\end{equation}
in which 
\begin{equation}
r = \sum_{i=1}^{N_X}r_i\,.
\end{equation}
For simplicity, we present results for several values of the parameter
$r$ in our phenomenological analysis below.

There would be no essential difference in the couplings of $h$ and $H$ with vector-like quarks, gauge bosons, in Eqs. (\ref{hHyq}), (\ref{hHvc}),  (\ref{hHX1}),  (\ref{hHX2}),  (\ref{hHX12}), if other choices of the hypercharges for the  multiplets in Eq. (\ref{qvmult}) were done.
For example, we could have chosen hypercharges $7/6$ and $2/3$, $-5/6$ and $-1/3$, or even $-5/6$ and $-4/3$ for the doublet and singlet, respectively. One difference with these other choices  would be that the effective interaction of $H$ with the electromagnetic field might be modified due the dependence on the the electric charge of the vector-like
quarks, which affects the branching ratio for the decay into two photons,
$BR(H\rightarrow\gamma\gamma)$. In the scenario we consider, the main decay channels of $H$ are the ones involving a pair of massive vector gauge bosons, $V=Z,\,W$, with corresponding branching ratios are such that
$BR(H\rightarrow V\,V)\gg BR(H\rightarrow\gamma\gamma)$. In fact, the decay channel into two photons in our case have $BR(H\rightarrow\gamma\gamma)< 0.001$ and by this reason it is not considered in our analysis for $H$ production signals.
Under this same assumptions the results of our analysis on the heavy Higgs can be translated to similar models having other choices for the multiplets in Eq. (\ref{sdfds}).

We now present two phenomenological analysis of the heavy Higgs: one for the observed diboson excess by the ATLAS Collaboration and the prospects to discover the new scalar boson at the 14 TeV LHC.

\section{Heavy Higgs Phenomenology}
\label{sec:pheno}
\subsection{Production of a heavy Higgs boson with vector-like quarks couplings}

The heavy Higgs is produced predominantly via gluon fusion when contributions from vector-like quarks couplings come into play. Weak boson fusion and Higgs-strahlung processes are also possible.

\begin{figure}[t]
\centering
 \includegraphics[scale=0.46]{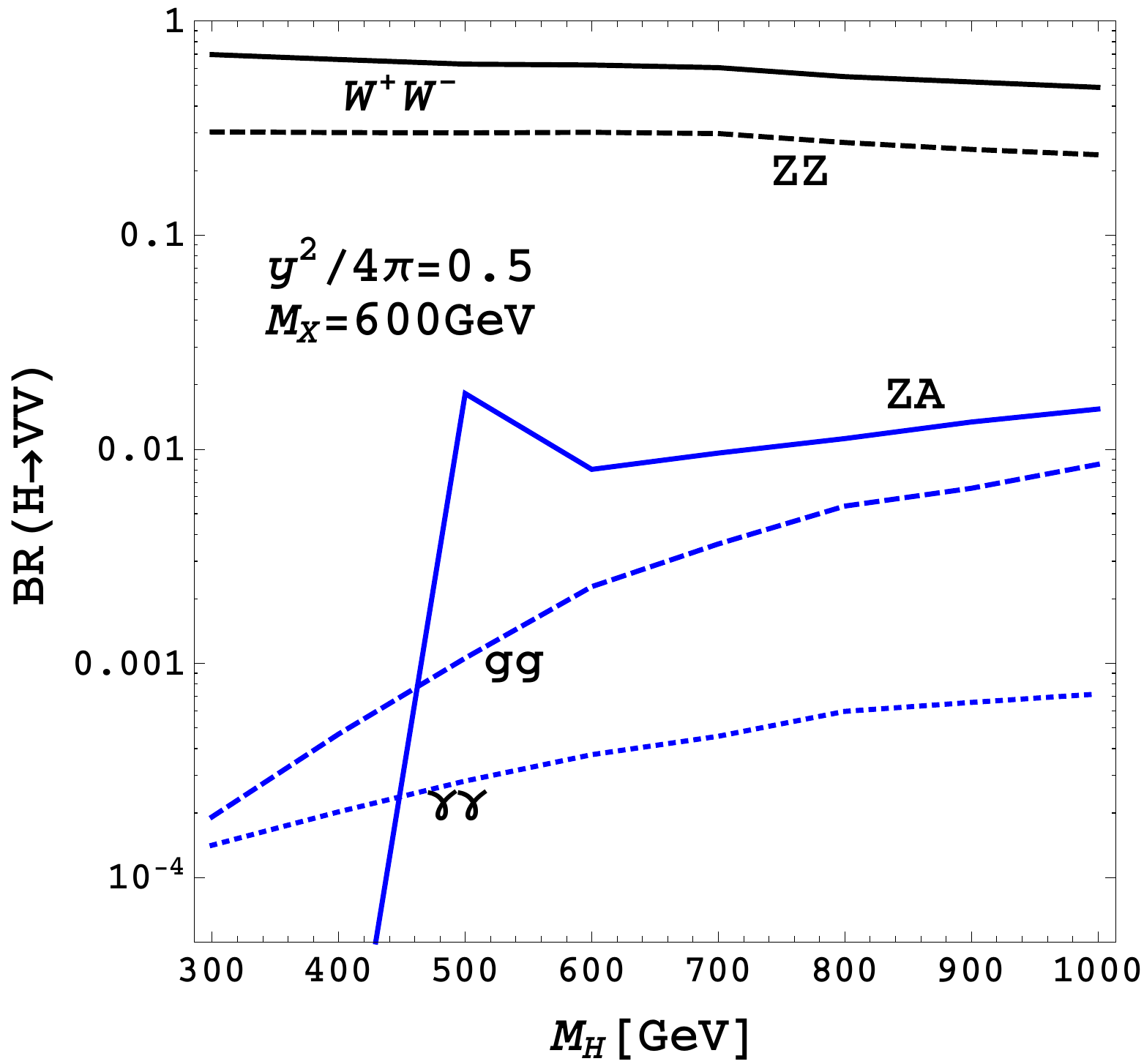}
  \includegraphics[scale=0.45]{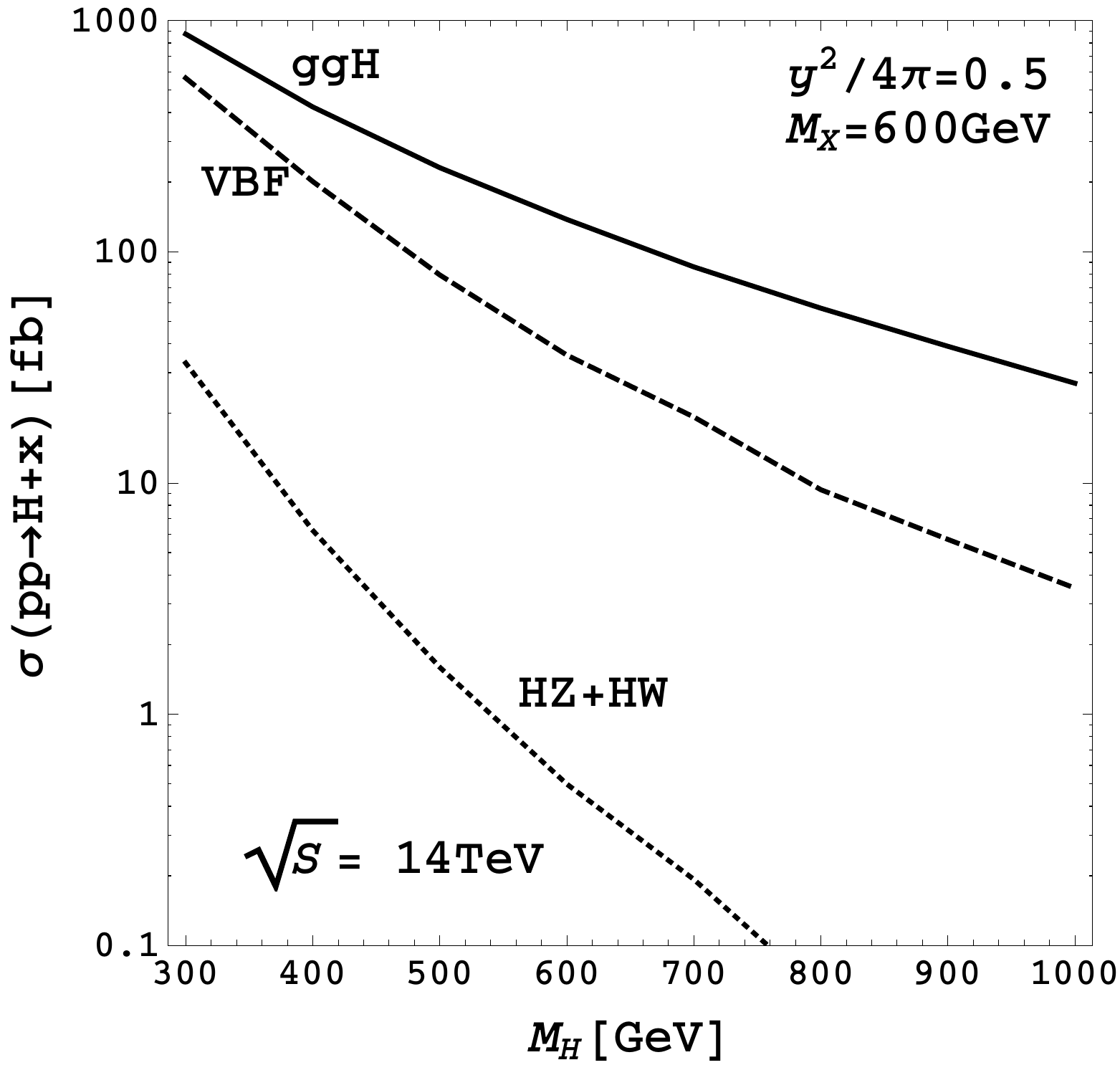}
\caption{Dominant branching-ratios (left panel) and cross section production rates of $H$ at the 14 TeV (right panel): WBF, the weak boson fusion; ggH, the gluon fusion; $HZ+HW$, Higgs-weak boson associated production (Higgs-strahlung). We fix $y=\sqrt{2\pi}$, $M_X=600$ GeV, and $m_{h^\pm}> m_H$ in these plots within the one doublet plus one singlet scenario as our base point.}
\label{xsecs}
\end{figure}

At the right panel of Fig.~(\ref{xsecs}) we show the $H$ production cross section at the 14 TeV LHC in the gluon fusion (ggH),  weak boson fusion (WBF), and $HV$, $V=W,Z$ associated production in the scenario with one doublet plus a singlet (one replica). The gluon fusion process is sensitive to the vector-like quark masses which run in the effective loop-induced gluon-gluon-Higgs coupling. The cross section $\sigma(gg\rightarrow H)$ is proportional to $(y^{ii}_{XH}v /M_{X_i})^2$. The WBF and $HV$ production cross sections, by their turn, are proportional to $c^2_{\alpha-\beta}$. The inclusive cross sections were computed at leading order with \texttt{MadGraph5}~\cite{Alwall:2014hca}.

\begin{figure}[t]
\centering
 \includegraphics[scale=0.40]{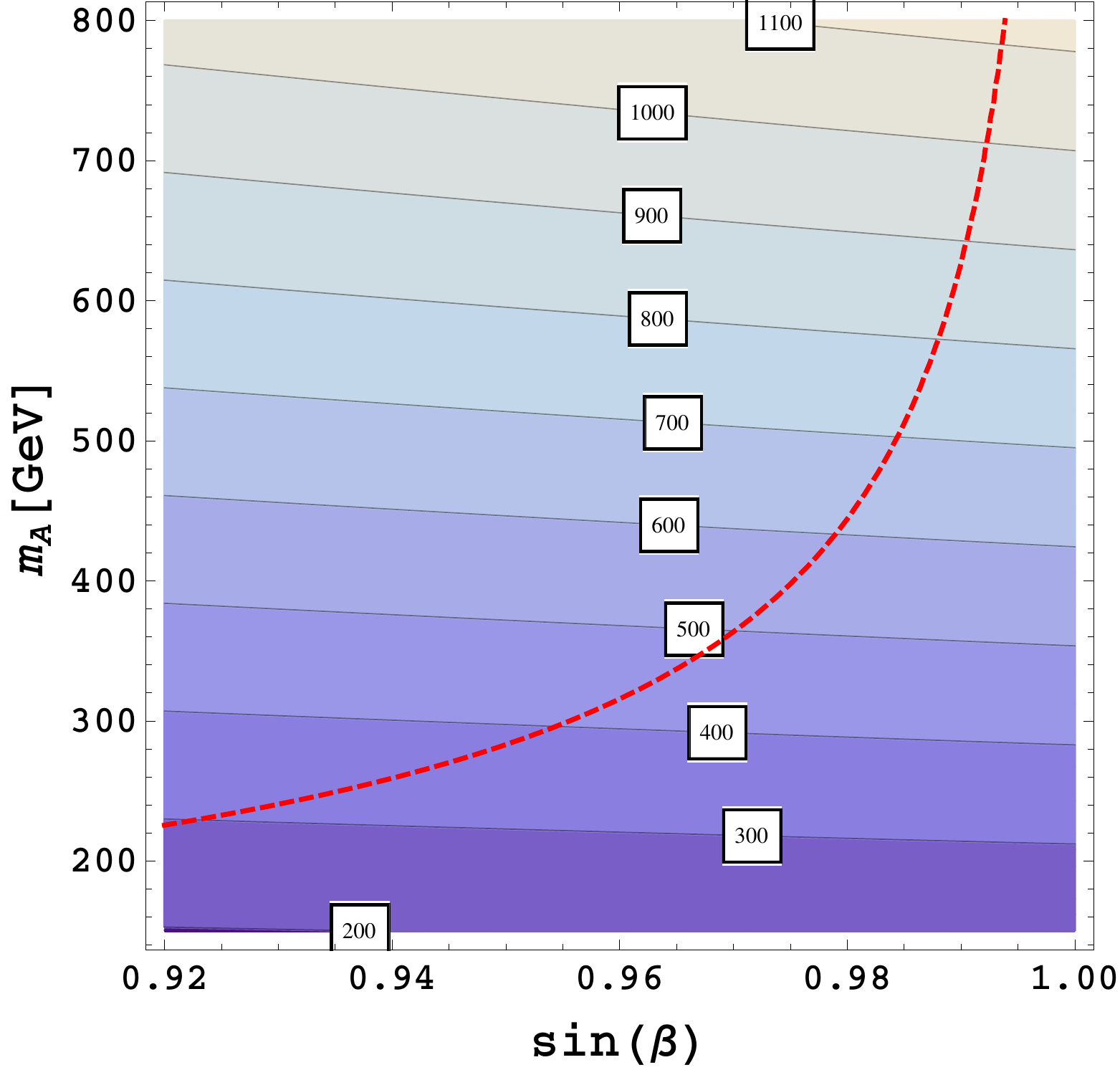}
 \includegraphics[scale=0.39]{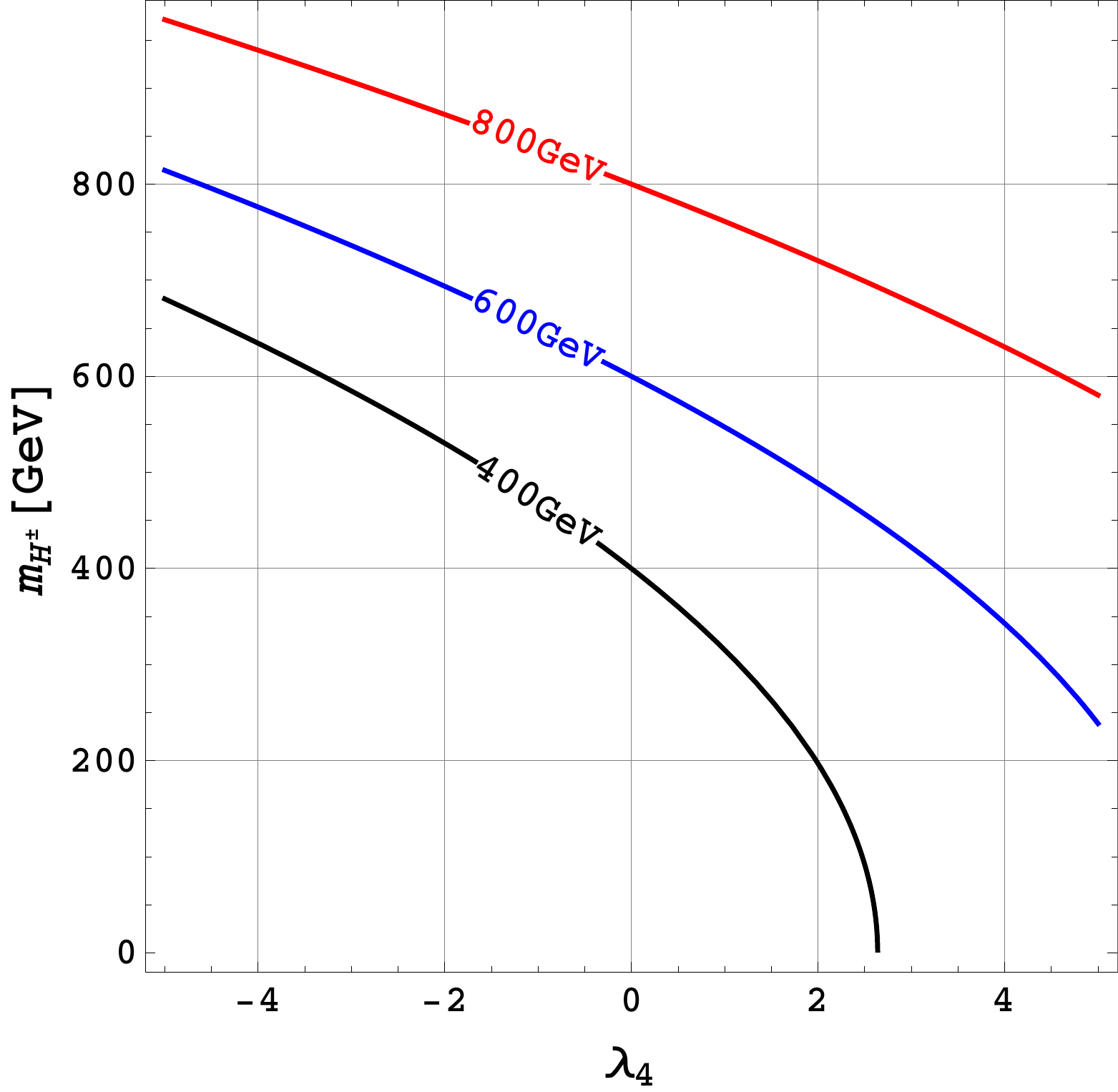}
\caption{At the left panel, the points on the red dashed line are those points where $m_h=125$ GeV in $s_\beta$ {\it versus} $m_A$ plane. The corresponding heavy Higgs masses are given by the nearly horizontal black lines in the plot. At the right panel, we can read the charged Higgs boson mass, as a function of the $\lambda_4$ parameter. Each line in the plot has the pseudoscalar mass $m_A$ fixed as 400, 600 and 800 GeV. For both plots we fixed $\lambda_1=0.01$, $\lambda_2=0.001$ and $\alpha=0$.}
\label{masses}
\end{figure}

Once we have chosen $\lambda_1,\; \lambda_2,\; \lambda_4$ and $m_A$, and given the constraint of $m_h=125$ GeV, the masses of the Higgs bosons $H$ and $h^\pm$, and $c_{\alpha-\beta}$ get determined from Eq. (\ref{mhH}). In Fig.~(\ref{masses}), at the left panel, the points on the dashed red line  satisfy $m_h=125$ GeV for $\lambda_1=0.01$ and $\lambda_2=0.001$, assuming $\alpha=0$. That is it, fixing a $H$ mass, there is only one pair $(s_\beta,\,m_A)$ consistent with the SM Higgs boson mass. We took this relation for all of our results. At the right panel, the lines represent points with $m_A$ fixed. This time, its possible to get a range of values for $m_{h^\pm}$ depending on $\lambda_4$.

The freedom in choosing the charged Higgs mass is important to keep the $T$ and $S$ parameters small, and mass splitting like $|m_H-m_{h^\pm}|<m_Z$ are desirable for this aim.  For example, with $m_A=600$ GeV, we see, from the left plot, that $m_H\approx 800$ GeV if $c_{\beta}<0.38$. Now, from the right plot, we see that a charged Higgs mass close to 800 GeV is possible for $-5\leq \lambda_4\leq -2$. Keeping $m_{h^\pm}$ close to $m_H$ is also important to close the decay channels of the neutral heavy Higgs into charged Higgs. We checked the values of  $T$ and $S$ for our the mass settings, taking the expressions of these parameters for the 2HDM, which are given in~\cite{Baak:2011ze}. We find consistence with the allowed values of $T$ and $S$ at 95$\%$ CL for all masses we use in our analysis.  For example: taking $m_A=300$ GeV, $s_\beta\approx 0.96$, $m_{h^\pm}=400$ GeV, we have $m_H\approx 411$ GeV which gives $T\approx -0.012$ and $S\approx -0.0011$; taking $m_A=500$ GeV, $s_\beta\approx 0.98$, $m_{h^\pm}=700$ GeV, we have $m_H\approx 697$ GeV which gives $T\approx -0.019$ and $S\approx 0.017$. 

Moreover, for small $\alpha$, the SM Higgs coupling to gluons respects the experimental bounds of~\cite{Khachatryan:2014jba} which, by the way, still allows for a 10\% deviation from the SM value.

Assuming all the Yukawa couplings $y^{ii}_{XH}\equiv y$ and $X$-quark masses equal $M_{X_i}\equiv M_X$, the ratio $r=\sum_i r_i$ is given by $r=N_X y\,v/M_X$ where $N_X=2$  for one new quark doublet and one singlet, that  is, for one single replica of quarks. We now seek the largest production cross sections permitted by the bounds to our parameters. Fixing $y^2/4\pi=0.5$ as the largest Yukawa coupling respecting perturbativity ($|y|<\sqrt{4\pi}$), we first require $c_{\alpha-\beta}\leq 0.38$ in order that our model predictions do not conflict with the current data. In Ref.~\cite{Chatrchyan:2013wfa}, a bound of approximately 800 GeV is obtained for vector-like $X$-quarks decaying 100\% to top plus a $W$ boson. That limit does not apply to our model once we do not assume that the new quarks decay predominantly to bottom or top quarks. In Ref.~\cite{Cacciapaglia:2012dd}, the more realistic case of non-negligible decays to the other families are taken into account to relax the experimental constraints and get better estimates of the bounds. It is shown that limits from 600--700 GeV should be expected in the more realistic case of $BR(X\rightarrow tW)<1$. Thus, we assume $M_X=600$ GeV for the one replica scenario $N_X=2$. In the next sections we will relax these parameters and evaluate the impact on the search reach of the LHC.



In the two replicas scenario $N_X=4$, for example, keeping $y=\sqrt{2\pi}$, we can raise the $X$-quarks masses to about 1.1 TeV in order to get the same production cross section of the one replica scenario for the gluon fusion. Note that if we fix $M_X=600$ GeV as in the one replica case, we can decrease the Yukawa couplings or $c_{\alpha-\beta}$ getting away of the limiting values for these parameters opening up a somewhat larger region of the parameter space to increase the cross sections. Of course, we can keep adding more quarks doublets and singlets to relax the Yukawa couplings and enhance the cross sections even further.

Assuming, for example, a set of values for the angles inside the region allowed by the constraints on the type-I 2HDM from the LHC~\cite{Craig:2013hca},~\cite{Dumont:2014wha} as $\alpha\approx 0$ and $\beta\geq 3\pi/8$ (which gives $c_{\alpha-\beta}\leq 0.38$), we see from Eq. (\ref{mhH}) and the left panel of Figure~(\ref{masses}) that $m_H\geq  1.3\,m_A$.

From now on we define as our {\underline{base model}}: \\
\begin{equation}
N_X=2, \;\; M_X=600\;\hbox{GeV}, \;\; m_{h^\pm}> m_H, \;\; c_{\alpha-\beta}=0.38\;\; \hbox{and} \;\; y=\sqrt{2\pi}\; .
\end{equation}

This base model is just the parameters we chose to generate the events used to evaluate the cut efficiencies and the vector of features necessary to further processing in our decision trees analysis. With $m_H$ fixed, the cut efficiencies and the shape of the kinematic distributions do not depend on any other parameters. The cross sections and branching ratios, on the other hand, are rescaled from the base model values for each relevant point of the parameter space.

For the base model, the gluon fusion process rate is around 100 fb for a $700$ GeV Higgs. Interestingly, WBF has nearly the same cross sections for all masses from 300 to 1000 GeV at $\sqrt{S}=14$ TeV. The sum of $HW$ and $HZ$ rates are much smaller for a 700 GeV Higgs, reaching 1 fb. As we discussed, with more replicas of vectorial quarks we can adjust $M_X$, $c_{\alpha-\beta}$ and $y$ in a way we have the same cross sections of the base model.

At the left plot of Fig.~(\ref{xsecs}) we show the dominant branching ratios of $H$ for the base point too. The leading channels are $WW$ and $ZZ$ decays, followed by $ZA$; however, this later channel represents just about one percent of the $H$ decays for $400\, \hbox{GeV}<m_H<1$ TeV. The decay channels $\gamma\gamma, hh, Z\gamma$ and $gg$ represent less than 1\% of the decays in the the entire mass range up to 1 TeV as we see in Fig.~(\ref{xsecs}). The $WW$ and $ZZ$ branching fractions are weakly dependent on $c_{\alpha-\beta}$ in the region of the parameter space where they dominate the Higgs total width. Decays to photons and gluons depend mainly upon the ratio $r$. Here and in the forthcoming discussions we assume that $M_H<2M_X$, otherwise the new quarks would dominate the Higgs decays.

We also verified that the above base point is safe from the stringent bounds from heavy Higgs searches of the LHC@8 TeV runs in the channels $H\rightarrow b\bar{b}$~\cite{Khachatryan:2015yea}, $H\rightarrow hh$~\cite{Aad:2014yja}, $H\rightarrow ZZ,WW$~\cite{Khachatryan:2015cwa}, and $H\rightarrow \gamma\gamma$~\cite{Khachatryan:2015qba,Aad:2014ioa} for Higgs masses up to 1 TeV. As the cross sections decrease with $\cos(\alpha-\beta)<0.38$, all the points simulated further are also permitted.

In Sec.~\ref{multiv} we present the search analysis for the heavy Higgs at the 14 TeV LHC -- first, let us discuss about the diboson excess observed by the ATLAS Collaboration at the 8 TeV LHC.

\subsection{The diboson excess at the LHC 8 TeV}
\label{diboson}
The ATLAS Collaboration has recently reported an excess of events, as compared to the SM background, in a search for high-mass resonances in hadronically decaying weak bosons pairs $WW$, $ZZ$ and $ZW$~\cite{Aad:2015owa} amounting to $2.6\sigma$, $2.9\sigma$ and $3.4\sigma$, respectively. The CMS Collaboration also found an excess in the search for a high-mass resonance in the $(W\rightarrow \ell+\nu)+(h\rightarrow jets)$ channel with a global significance near $2\sigma$~\cite{Khachatryan:2014hpa,Khachatryan:2014gha}. Both studies involved jet substructure techniques to identify the weak and the SM Higgs bosons. The ATLAS search, for example, concentrated in the high-invariant mass region of jets pairs where a very heavy resonance decaying to weak bosons would produce two colimated ``fat jets'' from the bosons decays.

Recently, both ATLAS and CMS reported the first results from the Run II at 13 TeV and found no significant deviation above SM backgrounds~\cite{diboson13}. However, after combining the results from both experiments and comparing the experimental sensitivities of the Run I and II, the authors of Ref.~\cite{Dias:2015mhm} found that the preliminary results from Run II are still compatible, within around $1\sigma$, with the Run I results, even though the overall significance of the signal has dropped a bit. After all, the situation is blurred now, more data from the LHC at 13 TeV will be needed to confirm the excess.

In the ATLAS study~\cite{Aad:2015owa} it was further suggested that a new heavy $W$ boson with a mass $\sim 2$ TeV could fit the signals. Since then, several works appeared offering alternative explanations in terms of a scalar resonance for that excess~\cite{Aguilar-Saavedra:2015rna}--\cite{Sierra:2015zma}. In Ref.~\cite{Sierra:2015zma}, for example, it is shown that a Heavy Higgs coupling to vector-like quarks pertaining to higher $SU(3)_C$ representations presents a cross section in the 1--10 fb ballpark, the preferred region of the experimental study. The additional colors for the new quarks are essential to enhance the Higgs production cross section. A large number of vectorial quarks was the solution adopted in Ref.~\cite{Chen:2015cfa} to increase the cross sections. Moreover, an analysis of the invariant mass profile of the jets indicates that the resonance should be $\Gamma < 200$ GeV, which is also achieved in the models of Ref.~\cite{Chen:2015cfa,Sierra:2015zma}.

  In order to raise the production cross section of our 2 TeV Higgs boson, we  tried the type of solution proposed in Ref.~\cite{Chen:2015cfa} -- raising the number of vectorial quarks with masses larger than 1 TeV. In our approach, we keep the new quarks in the triplet representation of $SU(3)_C$ as the SM quarks.

\begin{figure}[t]
\centering
\includegraphics[scale=0.44]{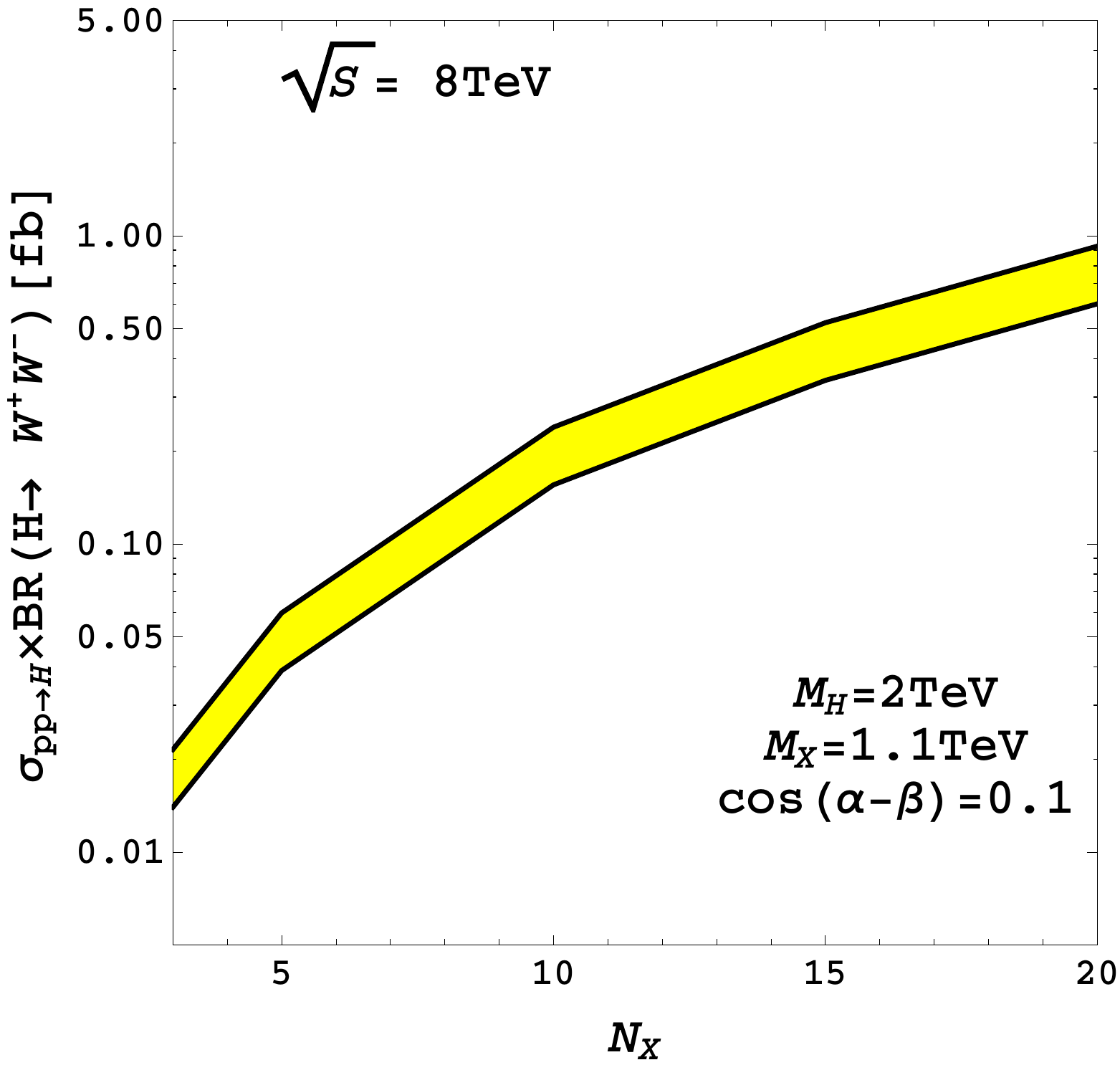}
\includegraphics[scale=0.44]{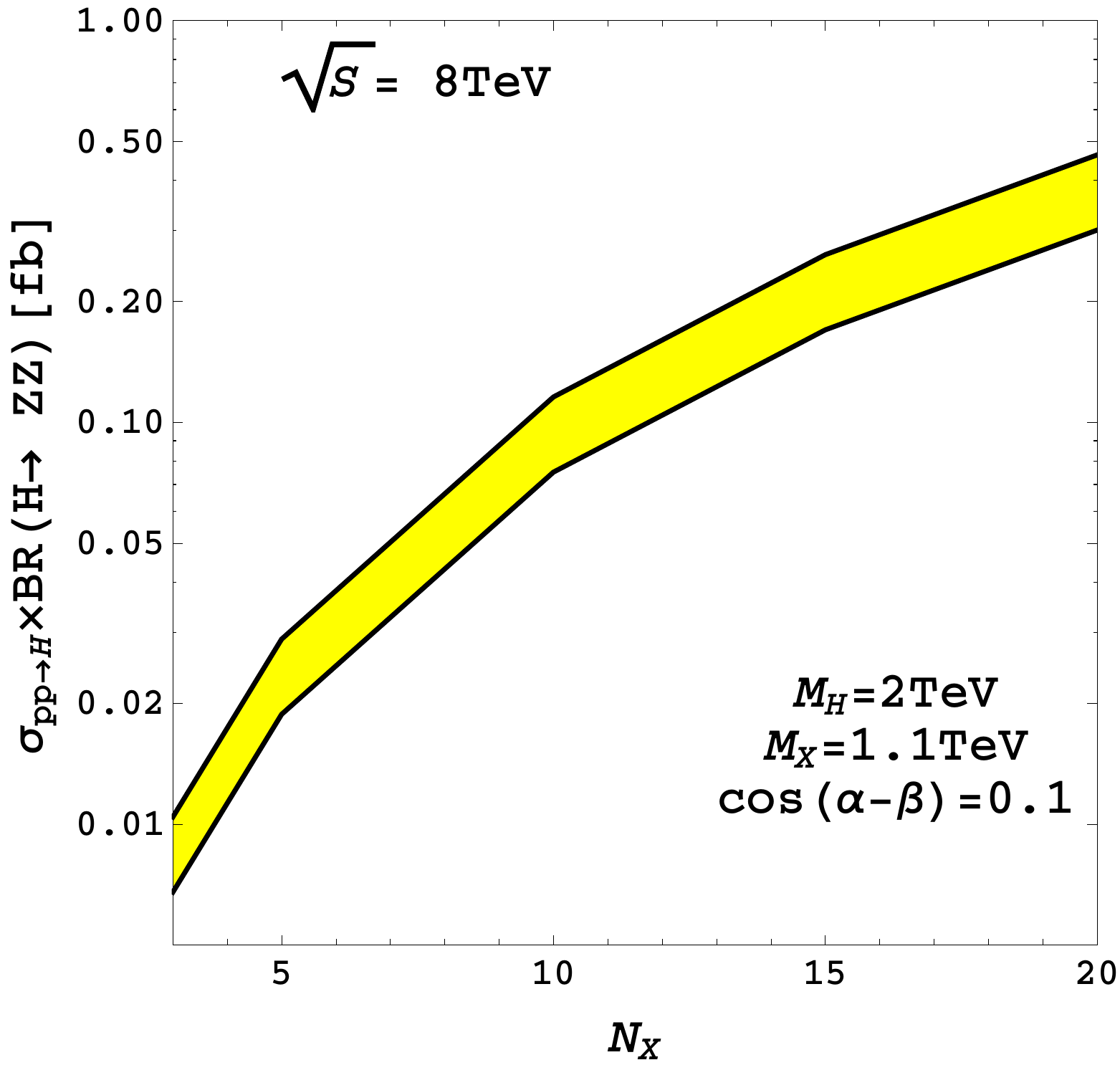}
\caption{In the left and right panels, we show the production cross section in gluon fusion times Higgs branching ratio to $WW$ and $ZZ$, respectively, in terms of the number of vector-like quarks at the LHC 8 TeV assuming $c_{\alpha-\beta}=0.1$. The Higgs mass is fixed in 2 TeV, the new quarks have masses of 1.1 TeV and $y=\sqrt{2\pi}$. The shaded bands represent and estimate of the uncertainty due the higher order QCD corrections to the production cross section.}
\label{excess}
\end{figure}

In the left and right panels of Fig.~(\ref{excess}), we show that in this model, only with a large number of new vector-type quarks, $N_X > 20$, we are able to reproduce a $WW$ or $ZZ$ excess, respectively, in the 1--10 fb region. We fixed the vectorial quark masses to 1.1 TeV in order they do not participate the decay of the heavy Higgs and do not contribute to the running of the strong coupling involved in the gluon-gluon-Higgs vertex, $y=\sqrt{2\pi}$ and $c_{\alpha-\beta}=0.1$, which is the maximum value permitted for a 2 TeV heavy Higgs in order that the Higgs boson mass gets fixed in 125 GeV in this model.

The shaded bands in Fig.~(\ref{excess}) correspond to an approximate QCD correction uncertainty. The lines limiting the shaded regions from below do not take any further K-factor correction, the cross sections are the matched ones given by \texttt{Pythia}+\texttt{MadGraph} with two extra jets which are around 30\% larger than the LO cross section for $2\leq N_X\leq 20$. The upper lines represent the simulated cross sections multiplied by an extra factor in order to reach the NNLO QCD correction to a heavy SM-like Higgs boson~\cite{Djouadi:2005gi} with a K-factor $\sim 2$. We are not aware of any NLO QCD correction to Higgs production in the gluon fusion process with several heavy vector-like quarks in the loop; thus, we adopt the heavy SM-like case as a first approximation.
A large statistical uncertainty is still involved in the determination of the production cross sections that can be inferred from the present ATLAS data set as stressed in~\cite{Allanach:2015hba}. In this work, a statistical analysis is performed to estimate the $ZZ$, $WW$ and $WZ$ cross sections which are compatible, within a certain confidence level, with the ATLAS data. It was found that large cross sections around 20 fb lie within the 95\% CL region, but also that the preferred rates are around 5 fb for $ZZ$ and $WW$ and a zero rate for the $WZ$. That is it, a neutral resonance decaying to $ZZ$ and $WW$ with similar cross sections at the 5 fb level might be the explanation of those observed excesses.

In order to have around 5 fb, we need around 50 new heavy quarks in the $WW$ channel and more than 50 in the $ZZ$ channel assuming $c_{\alpha-\beta}$. Putting aside the baroque aspect of a model with tens of new heavy quarks, it is not clear whether such high multiplicities would bring instabilities to the scalar potential. In the safe situation for the QCD asymptotic freedom with $N_X\leq 10$, however, the cross sections are too small to fit the excess if the heavy quarks are color triplets. Raising the $SU(3)_C$ representation, as in Ref.~\cite{Chen:2015cfa} would help to increase the cross sections, but we do not pursue this possibility in this work. Anyway, giving no excess at the 8 TeV LHC is by no means an excluded possibly in this moment.


From our results and those of~\cite{Chen:2015cfa,Sierra:2015zma} we conclude that a mechanism to enhance the Higgs production cross section is necessary if an explanation to the diboson excess in terms of a heavy Higgs coupling to new quarks is pursued in models of a heavy Higgs coupled to vector-like quarks. This mechanism could be, for example, a non-minimal Higgs sector or different $SU(2)_L$ representations for the new quarks and and/or Higgs bosons.

 It has to be pointed out that if a heavy Higgs is produced through a process involving a large number of extra quarks as we discussed, then, very probably, an additional mechanism to avoid destabilization of  the potential would also be required, for example by adding more scalars to the potential~\cite{EliasMiro:2012ay, Baek:2012uj, Lebedev:2012zw, Batell:2012zw}. The reason is that the new quarks give at first order a negative contribution to the beta functions of the scalar quartic couplings in the potential, potentially driving those couplings into negative values for large Yukawa coupling constants~\cite{Joglekar:2012vc}. Additional scalars, by their turn, contribute positively to the running of the quartic couplings, restoring the stability, at least up to a very high scale.

\subsection{Search analysis in the $H\rightarrow ZZ\rightarrow 4\ell$ channel}
\label{multiv}
Let us now explore the hypothesis of a heavy Higgs boson with mass up to 1 TeV, thus not associated with the ATLAS diboson excess. We want to estimate its discovery prospects at the LHC 14 TeV.

The large branching ratios to $ZZ$ and $WW$ make these channels obvious choices for a heavy Higgs hunting. In particular, the $ZZ$ channel allows us to reconstruct the heavy Higgs in a four charged leptons final state~\cite{Khachatryan:2015cwa}. A semi-leptonic channel with two jets and two charged leptons might also be interesting, but at the cost of higher backgrounds, contrary to the four charged leptons signal whose dominant background is the SM $Z$ pair production.

To show the potential of the LHC 14 TeV to discover a heavy Higgs boson that couples to vector-like quarks we simulate the gluon fusion process for our base model
\begin{equation}
pp\rightarrow H\rightarrow ZZ\rightarrow \ell^+\ell^-\ell^{\prime +}\ell^{\prime -}
\end{equation}
where $\ell$ and $\ell^\prime$ represent electrons or muons.

The heavy Higgs interactions were obtained with the help of \texttt{FeynRules}~\cite{Alwall:2014bza} and the partonic events were simulated with \texttt{MadGraph5}~\cite{Alwall:2014hca} . Two extra jets were also taken into account in the simulation in order to better estimate the kinematic distributions and cross sections. Hadronization and detector effects were taken into account from the \texttt{Pythia}~\cite{Sjostrand:2006za} and \texttt{Delphes}~\cite{deFavereau:2013fsa} interface to \texttt{MadGraph5}, respectively, within the $k_T$-MLM jet matching scheme~\cite{Mangano:2006rw}.

The relevant backgrounds for our signals are $ZZ$, $Z\gamma$, $\gamma\gamma$, $WWZ$, $t\bar{t}$, $b\bar{b}$ and $bbZ$ processes and were simulated with the same tools as used in the signal simulation. In order to suppress these backgrounds and select the candidate signal events we adopt the following basic cuts:
\begin{eqnarray}
& & p_{T}(1,2,3,4)>(20,20,15,15)\;\hbox{GeV}\; ,\; |\eta_\ell| < 2.5 \nonumber\\
70\;\hbox{GeV} &<& M_{\ell\ell} < 110\;\hbox{GeV}\; ,\; \etmiss < 40 \; \hbox{GeV}\; ,\; M_{4\ell} > 250\; \hbox{GeV}
\label{cuts}
\end{eqnarray}
where $p_T(n)$ denotes the $n$-th hardest lepton of the event.

The two hardest leptons are required to have a transverse momentum in excess of 20 GeV, while the two softest ones must have a transverse momentum of 15 GeV at least. All leptons are central with $|\eta_\ell|<2.5$. Same-flavor lepton pairs with invariant masses compatible with a $Z$ boson decay are selected by minimizing the variable $\chi^2_{4\ell}$ defined as follows,
\begin{equation}
\chi^2_{4\ell}=(M_{ij}-m_Z)^2+(M_{kl}-m_Z)^2\;\; ,\;\; i,j,k,l=\ell_{1,2,3,4}
\end{equation}
that is, by choosing the lepton pairs whose invariant masses are closest to the $Z$ boson mass. Once the $Z$ boson yields have been identified, we impose the $70\;\hbox{GeV}<M_{\ell\ell}<110\;\hbox{GeV}$ cuts.

We reject events with missing energy larger than 40 GeV to eliminate $WWZ$ and $t\bar{t}$ events. Backgrounds with bottom jets are efficiently cleaned up with lepton isolation criteria. Finally, a hard cut on the 4 leptons invariant mass $M_{4\ell}$ helps to identify typical leptons from a heavy resonance decay.

After those cuts, almost all reducible backgrounds are eliminated, but the irreducible $ZZ$ background remains much larger than the signal, amounting to $4.7$ fb. A 300 GeV Higgs, by its turn, has a cross section of $0.38$ fb after cuts for the base model. Hardening the $M_{4\ell}$ cut suppresses even further the $ZZ$ background in the search for a Higgs heavier than 300 GeV, but the production cross section for heavier Higgses decreases fast compensating for the background rejection and keeping the signal over background ratio at the $0.1$ level.

\begin{figure}[t]
\centering
 \includegraphics[scale=0.5]{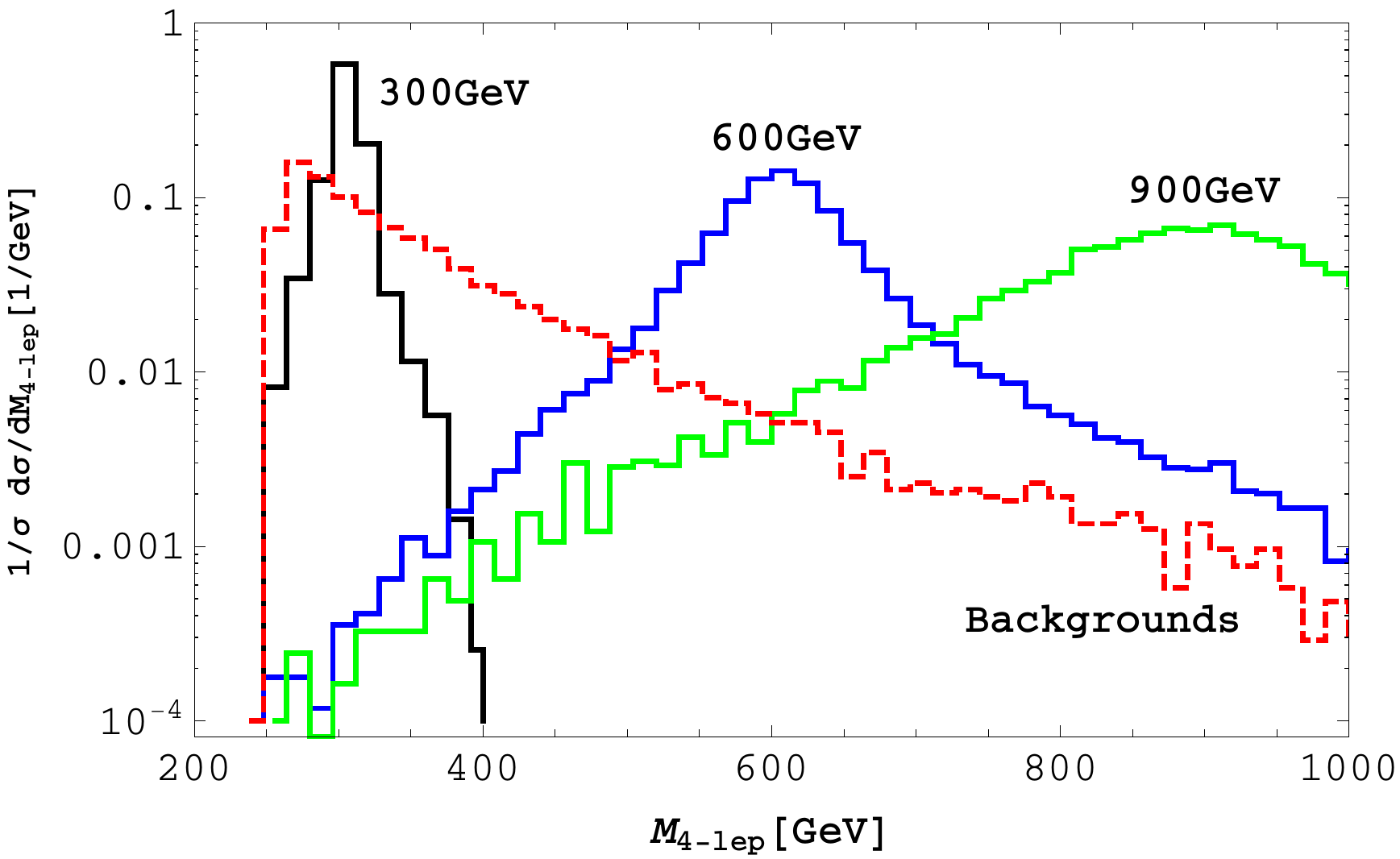}
\caption{The four-leptons invariant mass distribution for 300 GeV (black line), 600 GeV (red line) and 900 GeV (green line) Higgs bosons in the base model. The backgrounds are shown in the red dashed line.}
\label{m4lep}
\end{figure}

Cutting in a window around the Higgs mass does not improve the situation when the resonance gets broad as the mass increases, throwing away too much signal; see Fig.~(\ref{m4lep}). This is precisely the region where the $HZZ$ coupling is large. Moreover, a small $S/B$ ratio makes it impossible to discover a heavy Higgs when we take even modest systematic uncertainties into account. Thus, for most part of the parameter space where $H\rightarrow ZZ$ is a good searching channel, the broad resonances associated with the $H$ decays and the small cross sections prevents us from identifying the signal events with a straightforward look at invariant mass windows even without taking systematics into account.
We present next a much better strategy to select signal events.

\subsection{Multivariate analysis - Boosted Decision Trees}

Instead of hardening the cuts with modest expected gain in the signal significance, we can build a discriminant variable based on boosted decision trees~\cite{bdt,Roe:2004na}. A decision tree is a supervised machine learning algorithm aimed to classify real (and categorical) valued vectors -- our collider events, for example -- based on a collection of features of those events. The goal is to train the algorithm to identify signal and background events with a high performance assigning to each one an output score which reflects the chance of being correctly assigned to its true class. Decision trees have been used in high energy physics to separate signal and backgrounds events in difficult situations where usual cut analysis based on kinematic distributions are less efficient, for example, in Higgs decays to tau leptons at the LHC 8 TeV~\cite{Aad:2015vsa}, single top production at the Tevatron~\cite{Aaltonen:2010jr}, $t\bar{t}h$ production and coupling measurement at the LHC 14 TeV~\cite{Moretti:2015vaa} and high-level triggering at colliders~\cite{Gligorov:2012qt}.

\begin{figure}[t]
\centering
 \includegraphics[scale=0.5]{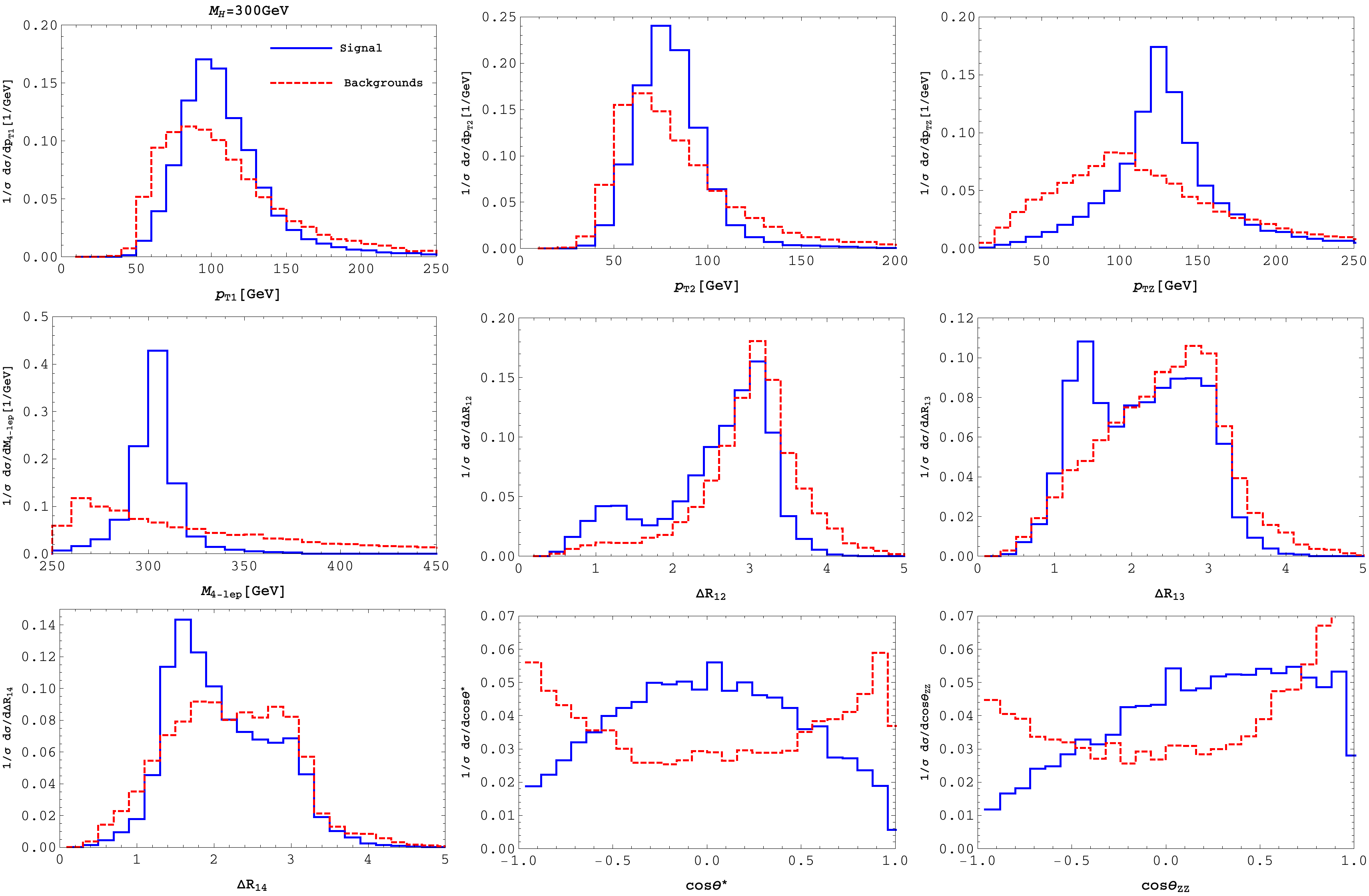}
\caption{The nine variables used to build the decision trees. The solid blue(red) lines display the normalized distribution of 300 GeV Higgs bosons(backgrounds). At the upper row, from left to right, respectively, the hardest lepton, the second hardest lepton and the hardest $Z$ boson transverse momentum. The four-lepton invariant mass is shown at the middle left panel. $\Delta R_{1k}$ represents the separation between the hardest lepton and the three ($k=2,3,4$) other softest leptons of an event. The lower row contains also the distribution of the cosine of the $\theta^*$ and $\theta_{ZZ}$ angles as defined in the text.}
\label{vars}
\end{figure}

Basically, a decision tree classifies an event from a sequence of binary choices. These choices are based on a given set of observable features in an exhaustive(greedy) way, that is, by searching for the feature and the separation cut of that feature which maximizes some measure of purity of signal events. Starting from a root node where all events lie, each binary choice splits a node of the tree into other two, if the number of signal(backgrounds) events is high enough, the right(left) node, for example, is called a signal(background) node. The branching process ceases either when there is no improvement in the signal purity when a node is split up or a minimum number of events in the node is reached, then the node is promoted to a leaf - the prediction of the decision tree is contained in their leaves which can be signal or background leaves depending on the signal purity. Counting the number of signal events in a background leaf and vice-versa one can evaluate the misclassification rate of the algorithm and devise ways to reduce it.

The algorithm is powerful in classifying events but is very sensitive to the training sample. Statistically different samples may cause large variations on the tree configuration and the results. In order to overcome this difficulty a {\it boosting} procedure may be used. The basic idea is to identify and penalize the wrong assignments in order to enhance the purity of the leaves. Each time this re-weighting procedure is applied a new tree is built. After building several trees, the weights of signal and backgrounds events in the leaves of each tree is collected and combined to compute the BDT output score of the events. A good separation assigns to most signal events scores near $+1$, while backgrounds are concentrated towards $-1$. After combining all features into a BDT discriminant  distribution, a cut is chosen in order to maximize the statistical significance of the signal.

In this work we use a version of an adaptive boosting algorithm which stabilizes and improves the performance of the decision trees~\texttt{ADABOOST.MH}~\cite{adaboost} as implemented in  the \texttt{C++} package \texttt{Multiboost}~\cite{multiboost}. The main improvement of this algorithm as compared to the original boosting proposed in~\cite{shapyre} is a number of methods to tune the parameters that control the training error in a fast and efficient way. We trained a forest of 3000 one-decision two-leaf trees with \texttt{Multiboost}, after imposing the cuts of Eq.~(\ref{cuts}), in order to separate events into two classes  -- the signal class and the dominant $ZZ$ background class based on nine kinematic distributions (features).

We show in Fig.~(\ref{vars}) the nine normalized observable distributions (features) used to build the decision trees and classify our signal and background events -- the transverse momentum of the two hardest leptons $p_{T_{1,2}}$, the separation $\Delta R_{ij}=\sqrt{\Delta\eta_{ij}^2+\Delta\phi_{ij}}$ in the $\eta$-$\phi$ space between the hardest lepton $i=1$ and the three softest leptons of the event $j=2,3,4$, the hardest $Z$ boson transverse momentum $p_{T_Z}$, the four-lepton invariant mass $M_{4\ell}$, the cosine of the production angle of the $Z$ bosons $\cos\theta^*$ in the Lab frame, and the cosine of the angle between the $Z$ bosons 3-momentum $\cos\theta_{ZZ}$ in the Lab frame. The signal events correspond to our base model with $M_H=300$ GeV.

The signal leptons and $Z$ bosons (solid blue lines) are naturally harder and more colimated compared to the backgrounds (dashed red lines). The cosine of the production angle of the signal $Z$ bosons is flat at the partonic level as it originates from the decay of a scalar resonance, but its shape is distorted by detector effects, yet it is a good discriminating variable as we can see in Fig.~(\ref{vars}). In Fig.~(\ref{m4lep}) we show the four-lepton invariant mass distribution of the backgrounds and the signals for a 300, 600, and 900 GeV Higgs boson. As we pointed out, the resonance gets wider and a simple window cut wastes too much events. On the other hand, the shape of the distributions are still distinctive enough to help the building of the decision trees.

\begin{figure}[t]
\centering
 \includegraphics[scale=0.5]{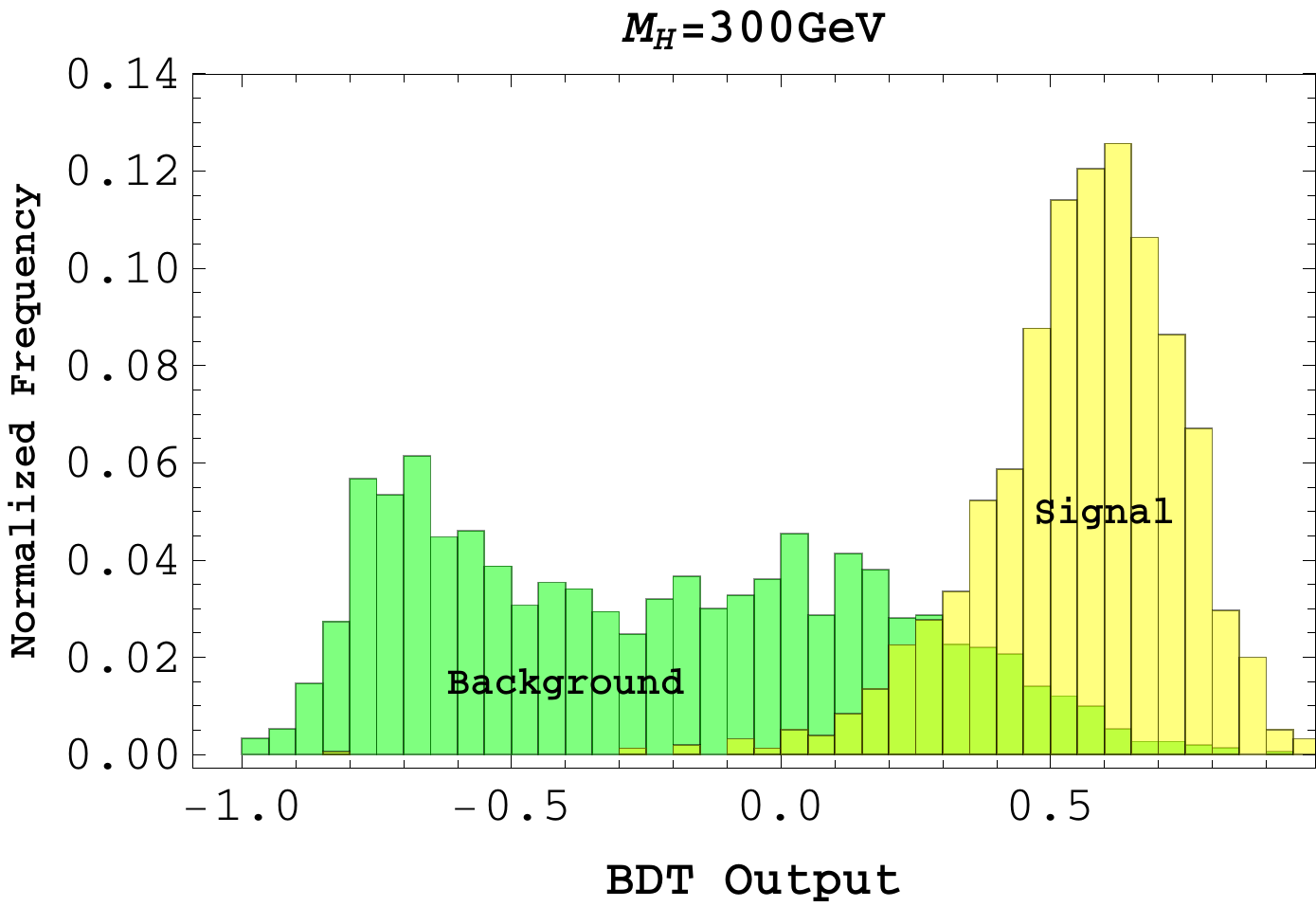}
 \includegraphics[scale=0.5]{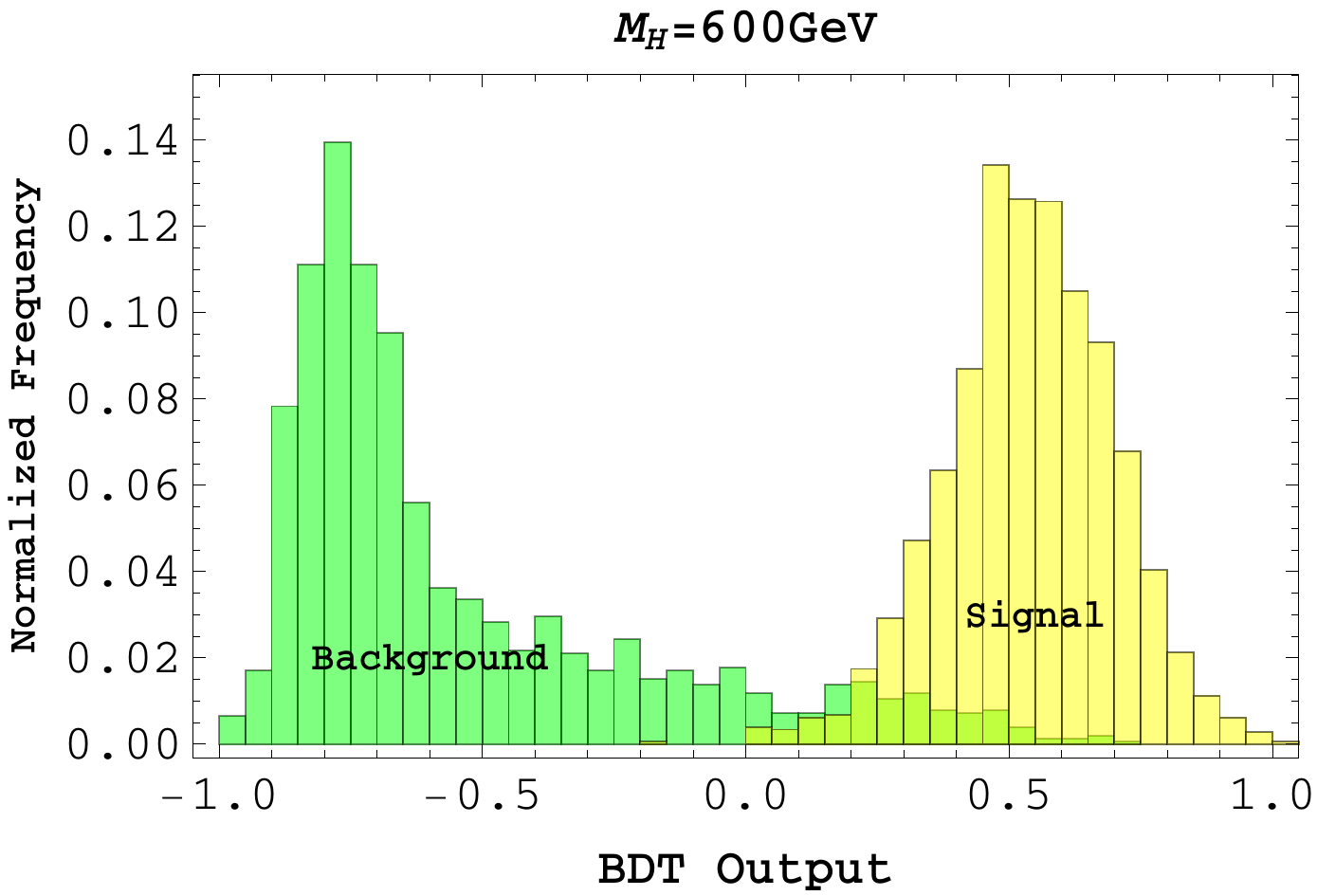}
\caption{BDT output distributions for signal and background events. At the left(right), we show the 300(600) GeV Higgs boson case versus backgrounds. Signal events are typically more strongly peaked towards the $+1$ score, while backgrounds in the opposite direction, near the $-1$ score. The dimensionful variables were decorrelated as explained in the text.}
\label{hists}
\end{figure}

After the cuts of Eq.~(\ref{cuts}), around $2\times 10^4$ events remain to train our decision forest. The simulated samples are split into 90\% for training and 10\% for validation. Typical BDT output distributions for the validation samples are shown in Fig.~(\ref{hists}) for signals and backgrounds. We see that the separation between the output distributions gets clearer as the Higgs boson becomes heavier. This is consequence of the more distinctive features of the heavier Higgs bosons compared to the backgrounds features. In fact, the chosen set of kinematic distributions turn the task of classification very efficient in this case. In special, note that some of the signal variables in Fig.~(\ref{hists}) have peak structures against smoother shapes of the backgrounds, for example, $p_{T_{ZZ}}$ and $M_{4\ell}$ -- events with peaking features are easily identified in BDTs enhancing its performance.

Despite an enhanced performance using the output distributions themselves is possible exploiting their shape information in a likelihood-ratio analysis as in Refs.~\cite{Aad:2015vsa,Aaltonen:2010jr,Cranmer:2015bka}, taking into account the systematic uncertainties is more complicated in this case. Instead, we just impose a cut on the BDT output to eliminate background events more efficiently. For example, imposing a BDT score larger than $0.5$, we see from Fig.~(\ref{hists}) that almost all backgrounds can be eliminated while keeping a good fraction of signal events.

The performance of a multivariate discriminant can be much improved when correlations among the variables are taken into account. In particular, a decision tree is an algorithm that is insensitive to the inclusion of irrelevant features to the discrimination. In this respect, not taking into account the correlations might turn a feature irrelevant making the BDT less optimal.

In our analysis, we found that the chosen distributions with mass dimension, that is, the transverse momentum and the invariant mass distributions, have high positive linear correlation coefficients. On the other hand, the correlations among the dimensionless features are small and not all positive. Moreover, the linear correlation between dimensionful and dimensionless features are also small.

\begin{figure}[t]
\centering
\includegraphics[scale=0.4]{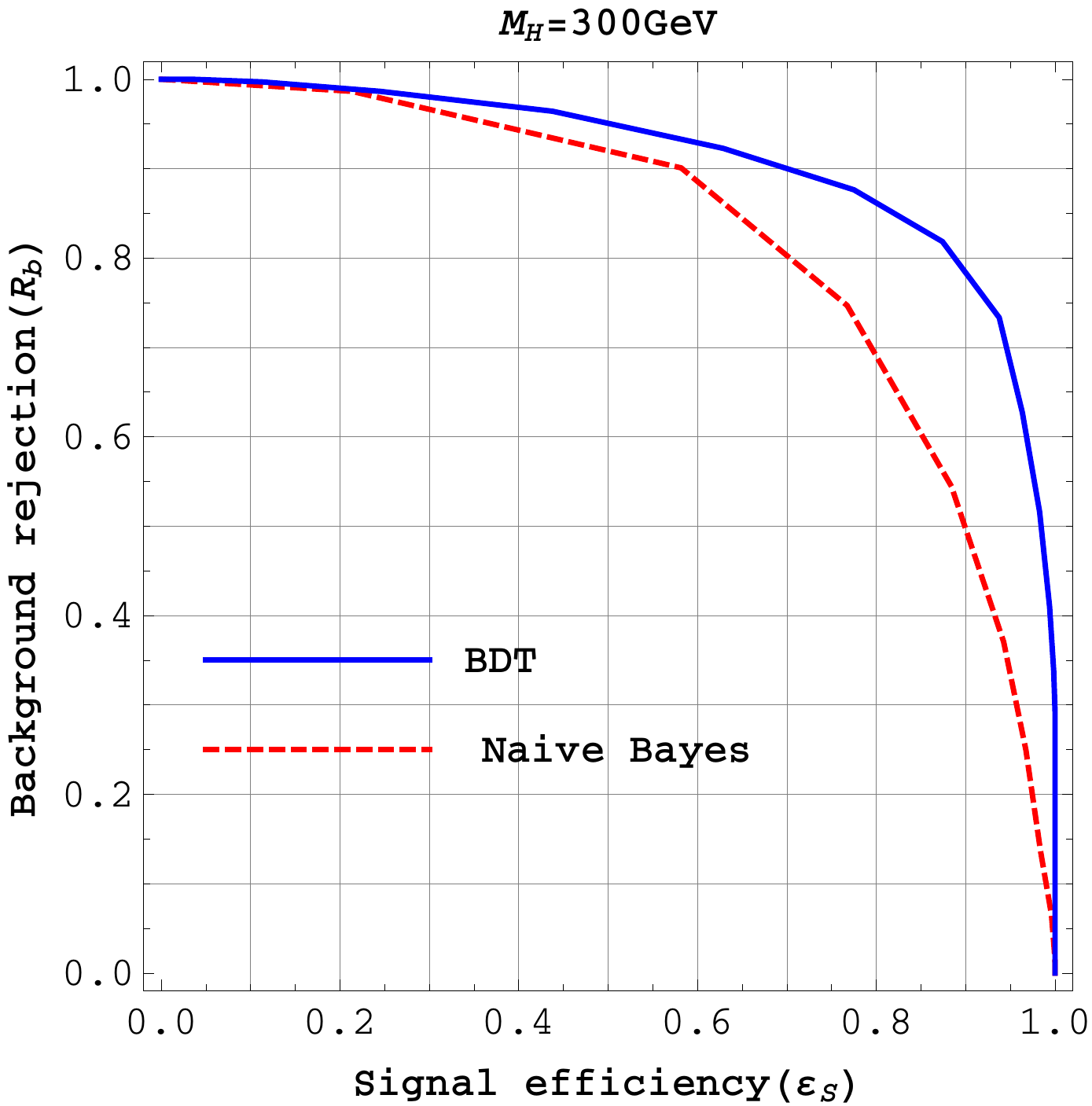}
\includegraphics[scale=0.4]{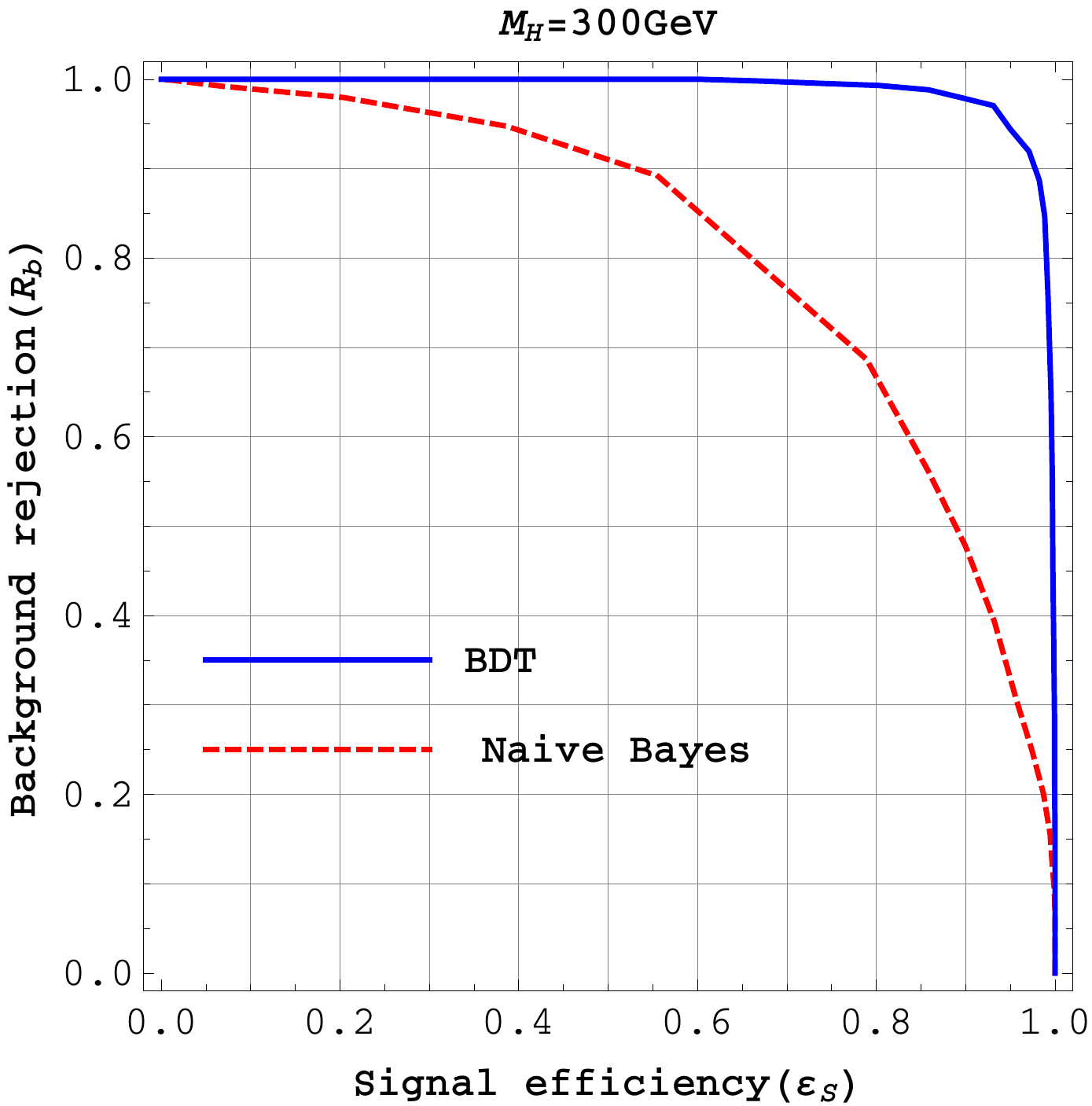}\\
\includegraphics[scale=0.4]{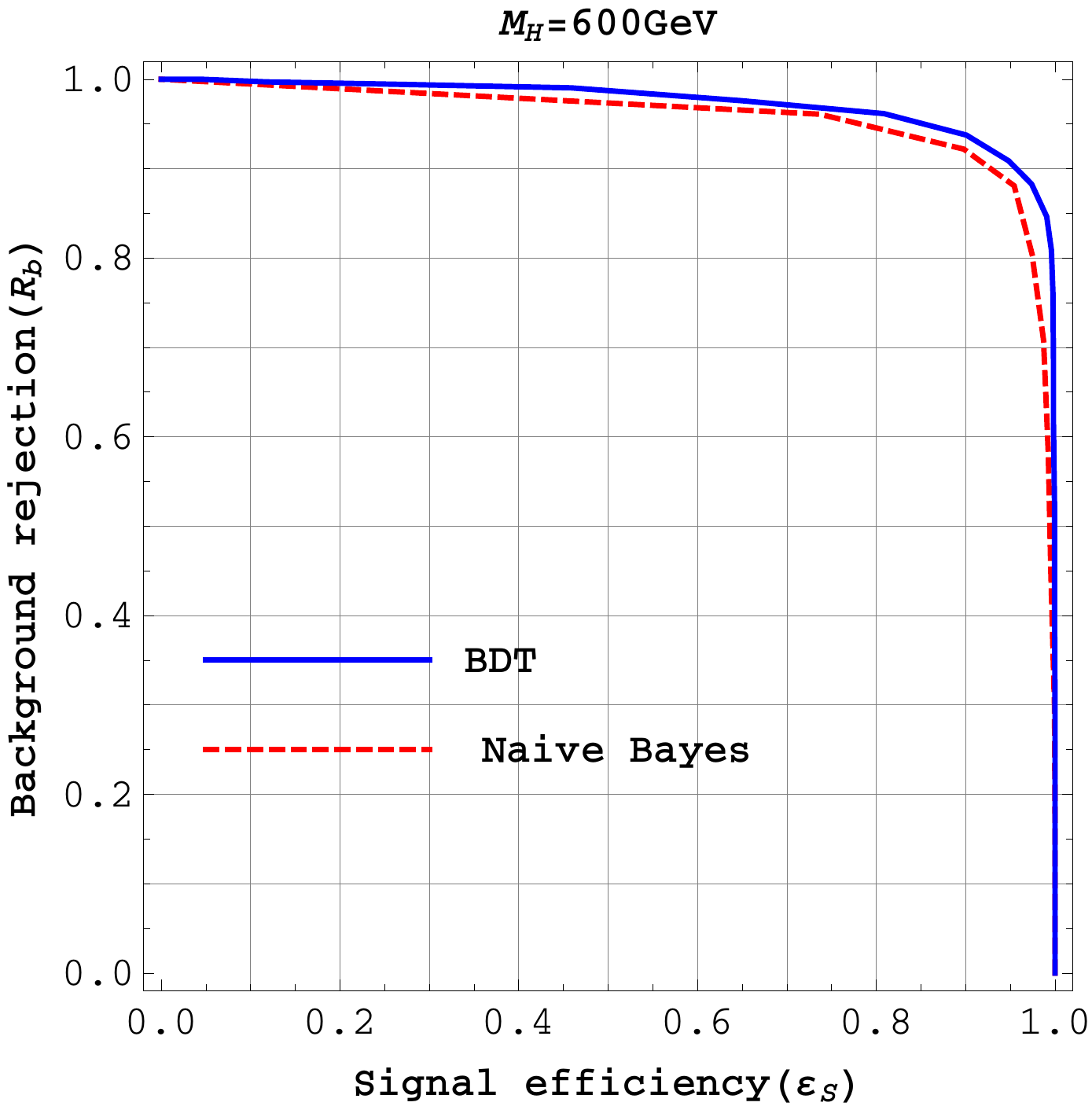}
\includegraphics[scale=0.4]{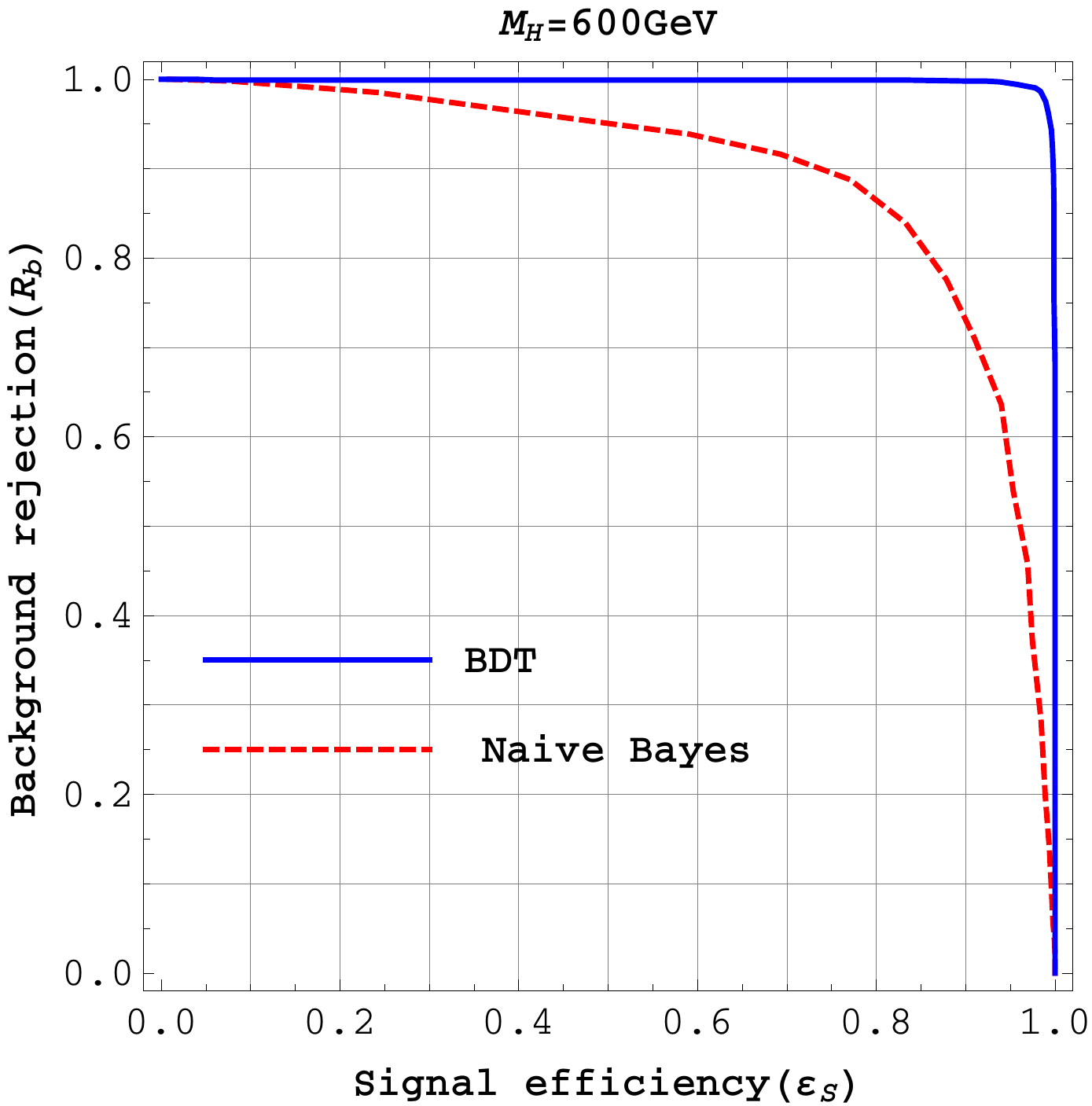}
\caption{Background rejection {\it versus} signal efficiency factors (ROC curves) from the BDT cut analysis are shown as blue solid lines. We show ROC curves for a 300 and a 600 GeV Higgs boson against backgrounds. The left panels display results with correlated dimensionful variables, and the right panels after decorrelation. The ROC curves for a naive Bayes discriminant are shown as dashed red.}
\label{roc}
\end{figure}

For this reason, we first decorrelate the dimensionful variables prior to their use to separate the events in the decision trees. This procedure indeed improves the performance of the discriminant. We can see this in Fig.~(\ref{roc}). In the left column, we show the background rejection against the signal efficiency (ROC curve) for nondecorrelated variables for a 300 and a 600 GeV Higgs at the upper and lower panel, respectively. In the right column,  we show the same but for decorrelated variables.

As the background rejection is typically high at the same time it keeps more than 90\% of the signal events, we choose to work at the fixed point of the ROC curves where the signal efficiency is 95\%. The background rejection achieved at this point after decorrelation is greater than 90\% for all Higgs boson masses and increasing from 95\% to almost 99\% from 300 GeV to 1 TeV Higgs bosons. The signal to background ratio is close to 10 for all these masses,  allowing a good discovery reach even for somewhat large systematic uncertainties in the background normalization. All the systematics analysis carried out in this work takes into account only the uncertainties in the background normalization.

We must point out again that the very good performance of BDT in the present case is the result of the several distinctive features used to distinguish between signal and backgrounds after decorrelating part of the variables as we discussed. We checked that the curves of the weighted error rates as a function of the boosting iterations (number of tress used in the forest) are monotonously decreasing; that is, there is no over-training.

The metric used to compute the signal significance which embodies systematic uncertainties in the number of background events is a profile likelihood-derived formula~\cite{LiMa}. It has been shown that this significance metric is superior to the usual naive metrics like $S/\sqrt{S+B}$ once it does not overestimate the significance in those regimes where the Gaussian approximation to the true Poisson distribution of signal and background events is not good~\cite{cousins}.

\begin{figure}[t]
\centering
 \includegraphics[scale=0.45]{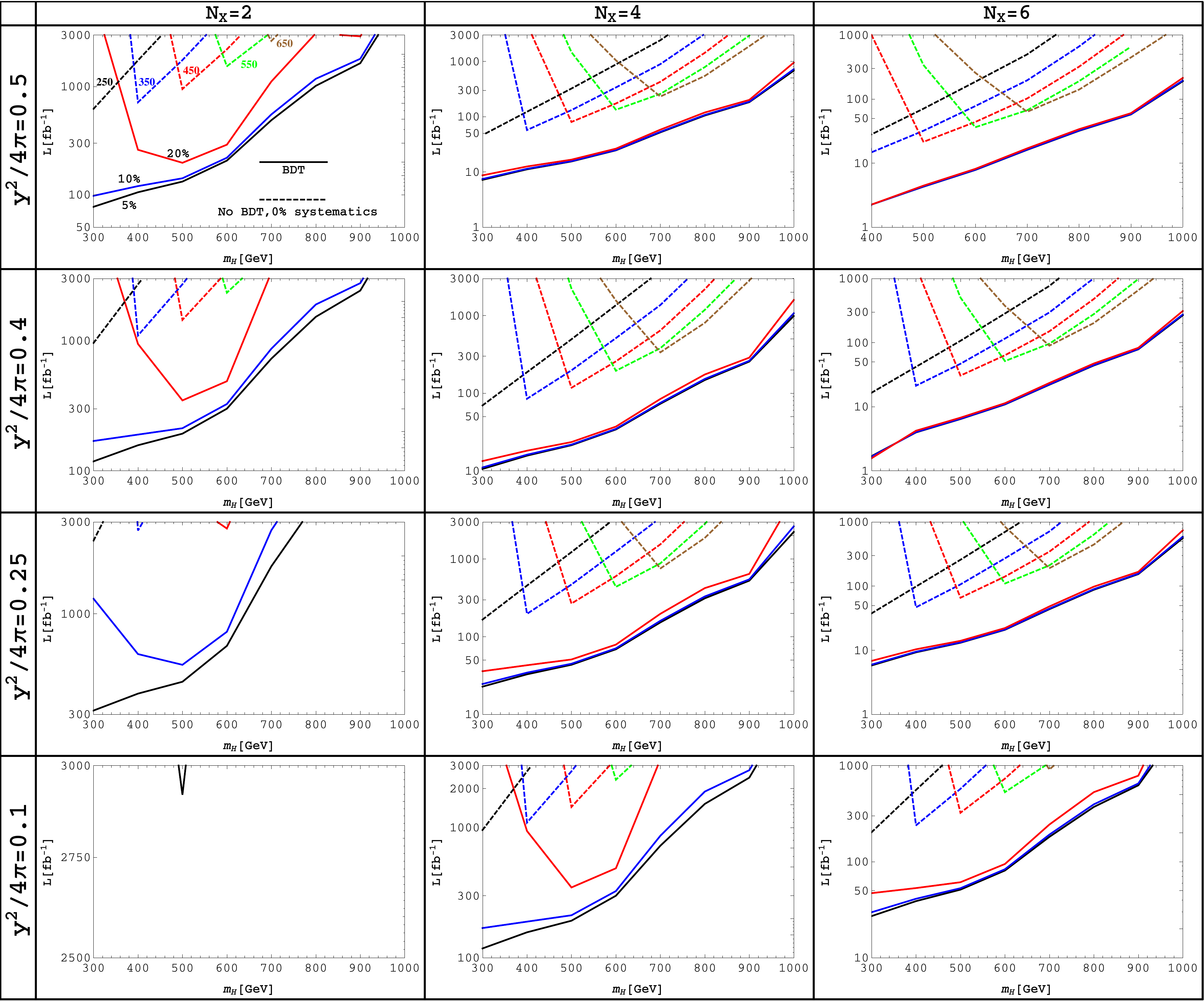}
\caption{Discovery reach of the LHC 14 TeV for the heavy Higgs boson as a function of its mass. The BDT analysis resulted in the solid lines assuming 5\%(black), 10\%(blue), and 20\%(red) systematics in the background normalization. We show twelve different points of the $\frac{y}{\sqrt{4\pi}}$ {\it versus} $N_X$ space. The dashed lines at the top of the plots are the discovery prospects based on hard $M_{4\ell}$ cuts only in an ideal situation with 0\% systematics.}
\label{results}
\end{figure}

In Fig.~(\ref{results}), we show the discovery reach for a heavy Higgs boson coupling with vector-like quarks as a function of the Higgs mass assuming nine different scenarios, from small $c_{\alpha-\beta}=0.05$ to the maximum presently permitted by the collider exclusions $c_{\alpha-\beta}=0.38$, and for three different Yukawa couplings $\frac{y}{\sqrt{4\pi}}=0.25$, 0.4 and 0.5, with and without the help of the BDT discriminant.

The upper dashed lines in Fig.~(\ref{results}) represent the $5\sigma$ discovery reach based on hard $M_{4\ell}$ cuts, namely, for the curves from left to right, $M_{4\ell}>250,350,450,550,650$ GeV, respectively. Each of these cuts maximizes the reach for a Higgs boson whose mass is larger but closer to the invariant mass cut. We assume no systematic uncertainties in this analysis. We checked that at a 5\% systematics level no discovery is possible with 3 ab$^{-1}$.

Cutting on the BDT output, a very efficient veto of background events is achievable with little loss of signal events as we discussed right above. The solid curves of Fig.~(\ref{results}) display the $5\sigma$ discovery region of the BDT analysis. Even for a 20\% systematics a heavy Higgs boson discovery is possible. Comparing the lower solid 5\% curve with the upper dashed ones, we realize the large gain from the univariate to the multivariate analysis.

Higgs decays to $WW$ and $ZZ$ are the dominant ones as we see in Fig.~(\ref{xsecs}). As a consequence, the branching ratio to $ZZ$ and the reaches vary only mildly with $c_{\alpha-\beta}$ as we see in the rows of Fig.~(\ref{results}).

Keeping the systematic uncertainties below around 10\%, it is possible to discover Higgs bosons with masses up to 700 GeV with 3 ab$^{-1}$ of integrated luminosity for all scenarios studied but that with $N_X=2$ and small Yukawa couplings $\frac{y}{\sqrt{4\pi}}=0.1$. Larger systematics deplete somewhat the reach as we see from the solid red lines. We point out once more that even very small systematic uncertainties preclude a $5\sigma$ discovery without the BDT classification of the events. Relying on a multivariate analysis is crucial for the discovery prospects in this case.

The columns of Fig.~(\ref{results}) display the discovery reaches holding the number of new vector-quarks $N_X$ fixed and varying the Yukawa coupling $\frac{y^2}{4\pi}$ from $0.1$ to $0.5$. Only for the one doublet scenario, $N_X=2$, and $\frac{y^2}{4\pi}=0.1$, the LHC would not be able to discover this Higgs boson even with BDTs. On the other hand, for two or three doublets of new vector-like quarks, Higgs masses up to 1 TeV, at least, can be probed, especially if the systematic uncertainties can be controlled to within 10\%---20\%. If $N_X=6$ and $m_H\approx 400$--600 GeV, the discovery is possible for all the Yukawa couplings we assumed with rather modest luminosities, around 10 fb$^{-1}$.

We must emphasize that these estimates are conservative in the regime of small $c_{\alpha-\beta}$. To obtain those results, we just rescaled the production cross sections and branching ratios of the base point. However, the total width of the new Higgs is dominated by the $ZZ$ and $WW$ partial widths which are proportional to $c^2_{\alpha-\beta}$, so the total width decreases quadratically with this parameter rendering the resonance much narrower for small $c_{\alpha-\beta}$. We do not take this effect into account, but the discovery reach can only improve for narrow resonances once a somewhat larger efficiency is expected concerning the $M_{4\ell}$ cut. Also the BDT performance is expected to remain the same, at least with a narrow $M_{4\ell}$ distribution. Therefore, we conservatively keep the operating point of the ROC curves for all scenarios studied.

As a final remark, we checked that a 2 TeV Higgs boson presents a too faint signal in the $H\to ZZ\to 4\ell$ channel to be observed at the 14 TeV LHC in this model even for the multivariate analysis. However, the cross section times branching ratio is promising for the $WW$ and $ZZ$ channels reaching around 7.5 and 2.3 fb, respectively. In both estimates we assumed 30 vector-like quarks with 1.1TeV running in the $Hgg$ loop-induced coupling as in the 8 TeV LHC diboson analysis presented in Sec.~\ref{diboson}.

We also found that a multivariate analysis based on a naive Bayes classifier, which is a much easier discriminant to build, performs well to separate signals and backgrounds as shown in the dashed red curves of Fig.~(\ref{roc}), but inferior to the BDT discriminant in all instances studied in this work. Overall, we advocate the use of multivariate techniques in the search for wide resonances where cut-based analysis performs poorly.

\section{Conclusions}
\label{sec:conclusions}
We propose a model with vector-like quarks, two Higgs doublets plus a scalar singlet in addition to the SM fermions. The model involves an additional $U(1)_{PQ}$ symmetry to solve the strong CP problem. In the low energy regime we have a type-I 2HDM where the heavy CP-even Higgs couples predominantly to the vectorial quarks enhancing its production cross section in the gluon fusion process.

Discovering a heavy Higgs boson coupled to vector-like quarks at the 14 TeV LHC within the one or two new quark doublets scenarios studied in this work is somewhat challenging in $pp\to H\to ZZ\to \ell^+\ell^-\ell^{\prime +}\ell^{\prime -}$. The discovery reach of the LHC with the full data set of 3 ab$^{-1}$ extends to 1 TeV Higgses, at least, for $y=\sqrt{2\pi}$, $N_X\geq 2$ and to vectorial quark masses around the TeV scale assuming systematic uncertainties in the background rates up to 20\%. For smaller Yukawa couplings, using BDTs to classify signal and background events also allows us to discover heavy Higgs bosons with masses up to 1 TeV, especially for the case of two and three new vector-quark doublets. It is crucial for these prospects to use a multivariate analysis to have a high signal to background ratio to tame the systematics. With that aim, we trained a forest of decision trees in order to better classify signal and background events. The kinematic distributions of the four charged leptons in the Higgs decay to $Z$ bosons present very distinctive features compared to the dominant $ZZ$ background resulting in a high background rejection and high signal acceptance.

On the other hand, we found that a simple cut-and-count analysis based solely on kinematic variables, as the four leptons invariant mass, is doomed to fail. Even with rather modest systematics below the 5\% level, just cutting on the invariant mass is not enough for a $5\sigma$ discovery with 3 ab$^{-1}$ as the resonances expected from the model are somewhat large which depletes considerably the cut efficiency. Jet substructure techniques might help to boost those prospects but we do not pursue this possibility here.

However, these jet substructure techniques might already have revealed a 2 TeV heavy Higgs in the 8 TeV run. The ATLAS diboson excess might be explained by the production and decay of the heavy CP-even Higgs boson of the model taking into account the current experimental uncertainties as obtained in~\cite{Allanach:2015hba}. As in Refs.~\cite{Chen:2015cfa,Sierra:2015zma}, however, a mechanism to raise the production cross section is necessary. Instead of promoting the new quarks to a higher $SU(3)_C$ representation as in~\cite{Sierra:2015zma}, a larger number of new vector-like quarks as heavy as 1.1 TeV can contribute to the Higgs-gluons loop~\cite{Chen:2015cfa}. About 50 new quarks would be necessary for a 5 fb signal in the $WW$ channel and 2 fb in the $ZZ$ channel in the base point but at the cost of a wider resonance, $\Gamma_H/ m_H\sim 1$. A narrow resonance with $\Gamma_H/ m_H\lesssim 0.17$ is possible setting $c_{\alpha-\beta}\lesssim 0.1$, but a 5(2) fb for the $WW(ZZ)$ channel would demand even more quarks. For a small number of new quarks, as $N_X<10$, the model predicts a too small diboson signal; however, we remark that, with the new LHC 13 TeV data, the diboson signal weakened considerably.

 As a final remark, we point out that the model contains a heavy CP-odd Higgs boson that can play the role of the 750 GeV resonance which seems to be arising in the 13 TeV LHC data in both ATLAS and CMS. We plan to investigate that exciting possibility in the near future.

\bigskip{}

\textbf{Acknowledgments.} 

The authors acknowledges financial support
from the Brazilian agencies CNPq, under Grants No. 303094/2013-3
(A.G.D.), and No. 307098/2014-1 (A.A.), FAPESP, under Grants No. 2013/22079-8 (A.A. and A.G.D.), and CAPES (D.A.C.).


\end{document}